\documentclass[a4paper,11pt]{article}
\pdfoutput=1
\usepackage{geometry}
\usepackage{jcappub}

\usepackage[normalem]{ulem}
\usepackage{lmodern}
\usepackage{graphicx}
\usepackage{aas_macros}
\usepackage{amsmath, bm}	
\usepackage{amssymb}	
\usepackage[dvipsnames]{xcolor}     
\usepackage{caption}
\usepackage{float} 
\usepackage{ulem}
\usepackage{xspace}
\usepackage{fontawesome}
\usepackage{hyperref}
\usepackage{orcidlink}
\usepackage{hanging}
\usepackage{geometry}
\usepackage{xcolor}
\usepackage[T1]{fontenc}
\newcommand{\hMpc}{\,h\, \mathrm{Mpc}^{-1}}

\newcommand{\Mpch}{\,h^{-1} \mathrm{Mpc}}
\newcommand{\Gpch}{\,h^{-1} \mathrm{Gpc}}
\newcommand{\Msunh}{\,h^{-1}\mathrm{M}_\odot}

\newcommand{\tensor}[1]{\ensuremath{\mathsf{#1}}}

\newcommand{\omegac}{\omega_{\rm cdm}}
\newcommand{\omegab}{\omega_{\rm b}}
\newcommand{\Neff}{N_{\rm eff}}

\newcommand{\Sigmasm}{\Sigma_{\text{sm}}}
\newcommand{\beq}{\begin{eqnarray}}
\newcommand{\eeq}{\end{eqnarray}}

\newcommand{\vect}[1]{\boldsymbol{#1}}
\newcommand{\lcdm}{$\Lambda$CDM}

\newcommand{\abacussummit}{{\tt AbacusSummit}}

\newcommand{\patchy}{{\tt Patchy}}
\newcommand{\nseries}{{\tt Nseries}}

\def\thickhline{\noalign{\hrule height.14pt}}

\usepackage[nameinlink,noabbrev]{cleveref}
\crefname{equation}{Eq.}{Eqs.}
\crefname{section}{Section}{Sections}
\crefname{figure}{Fig.}{Figs.}
\crefname{table}{Table}{Tables}
\crefname{appendix}{Appendix}{Appendices}
\Crefname{figure}{Figure}{Figures}
\Crefname{equation}{Equation}{Equations}
\Crefname{section}{Section}{Sections}
\Crefname{table}{Table}{Tables}

\title{Modelling the BOSS void-galaxy cross-correlation function using a neural-network emulator}
\author[a,b,1]{Tristan S. Fraser,\note{Corresponding author}}
\author[a,b]{Enrique Paillas,}
\author[a,b,c]{Will J. Percival,}
\author[d]{Seshadri Nadathur,}
\author[e,f]{Sla{\dj}ana Radinović,}
\author[e]{and Hans A. Winther}
\affiliation[a]{Waterloo Centre for Astrophysics, University of Waterloo, 200 University Ave W, Waterloo, ON N2L 3G1, Canada }
\affiliation[b]{Department of Physics and Astronomy, University of Waterloo, 200 University Ave W, Waterloo, ON N2L 3G1, Canada}
\affiliation[c]{Perimeter Institute for Theoretical Physics, 31 Caroline St North, Waterloo, ON N2L 2Y5, Canada}
\affiliation[d]{Institute of Cosmology and Gravitation, University of Portsmouth, Burnaby Road, Portsmouth, PO1 3FX, United Kingdom }
\affiliation[e]{ Institute of Theoretical Astrophysics, University of Oslo, P.O. Box 1029 Blindern, N-0315 Oslo, Norway}
\affiliation[f]{Institute of Space Sciences (ICE, CSIC), Campus UAB, Carrer de Can Magrans, s/n, 08193 Barcelona, Spain}
\emailAdd{tsfraser@uwaterloo.ca}
\emailAdd{epaillas@uwaterloo.ca}
\emailAdd{will.percival@uwaterloo.ca}
\emailAdd{seshadri.nadathur@port.ac.uk}
\emailAdd{sladana.radinovic@astro.uio.no}
\emailAdd{hans.a.winther@gmail.com}
\abstract{We introduce an emulator-based method to model the cross-correlation between cosmological voids and galaxies. This allows us to model the effect of cosmology on void finding and on the shape of the void-galaxy cross-correlation function, improving on previous template-based methods. We train a neural network using the \abacussummit\ simulation suite and fit to data from the Sloan Digital Sky Survey Baryon Oscillation Spectroscopic Survey sample. We recover information on the growth of structure through redshift-space distortions (RSD), and the geometry of the Universe through the Alcock-Paczyński (AP) effect, measuring $\Omega_{\rm m} = 0.330\pm 0.020$ and $\sigma_8 = 0.777^{+0.047}_{-0.062}$ for a $\Lambda \rm{CDM}$ cosmology. Comparing to results from a template-based method, we find that fitting the shape of the void-galaxy cross-correlation function provides more information and leads to an improvement in constraining power. In contrast, we find that errors on the AP measurements were previously underestimated if void centres were assumed to have the same response to the AP effect as galaxies---a common simplification. Overall, we recover a $28\%$ reduction in errors for $\Omega_{\rm{m}}$ and similar errors on $\sigma_8$ with our new method. Given the statistical power of future surveys including DESI and Euclid, we expect the method presented to become the new baseline for the analysis of voids in these data.}
\begin{document}
\maketitle

\section{Introduction}\label{sec:intro}

At late times, dark energy \citep{Riess1998,Perlmutter1999} rapidly expands the large, low-density regions of the matter field largely devoid of galaxies, called voids. Sandwiched between these voids, the galaxy distribution is highly non-linear on small scales, such that the galaxy two-point correlation function and power spectrum are statistically incomplete. 
A plethora of alternative statistics have been developed in order to recover information from the non-linear structures. Higher-order statistics \cite{Matsubara_1997,Zhang_2005,Schmittfull_2015,Gil_Mar_n_2016,Slepian_2018,Gualdi2019,Philcox2021} and alternative clustering statistics including wavelet scatter transforms (WST) \cite{Valogiannis_2022,valogiannis2023precise}, density split \cite{Paillas_2021,paillas2023cosmological}, and marked correlation functions \cite{white2016marked,Skibba2006} have been used to successfully boost the information obtained in clustering at late times. Using cosmic voids directly has, in particular, been a longstanding example of alternative clustering statistics being applied to provide additional cosmological information from galaxy redshift surveys \cite[e.g.][]{Lavaux_2012,Hamaus_2016,Nadathur_2020, Aubert_2022,Radinovi__2023, Woodfinden_2022,Woodfinden_2023,Paz_2013,Nadathur_2019_CMASS,Nadathur_2018,Hamaus2020:2007.07895,Contarini_2023}. Different aspects of voids have previously been used to test different components of $\Lambda$CDM. These include: the measured ellipticity as a test of the local tidal field \cite{park2007void}, the void density profile \cite{Paz_2013,Hamaus_2016,Nadathur_2017,Hamaus2014, yang2015} and the void-size function  \cite{Pisani2015,Sahlen_2016,Nadathur_VSF_2016,Contarini_2023,Massara_2015, Nadathur_2015}. The majority of recent analyses have focused on using void-galaxy cross-correlations \cite{Paz_2013,Hamaus_2016,Hamaus2020:2007.07895,Hamaus_2022,Contarini_2023,Woodfinden_2022,Woodfinden_2023,Nadathur_2018,Aubert_2022,Nadathur_2019_CMASS} to measure the geometry of the Universe through the Alcock--Paczyński (AP) effect \cite{AlcockPaczynski1979} and the growth of structure through redshift-space distortions (RSD) \cite{Kaiser1987} caused by the velocity fields around voids (although see \cite{Massara_2022}). This work aims to extend that success, developing a new method to model the void-galaxy cross correlations using a simulation-based emulator. This enables us to take into account several nuances in the way that voids are selected that are hard to model using template based approaches \cite[e.g.][]{Nadathur2019:1805.09349,Woodfinden_2022}. 

Measuring the clustering of galaxies around voids allows for the measurement of the growth rate of structure through $f(z)\sigma_8(z)$, which controls the amplitude of the RSD, and $D_{\rm M}(z)/D_{\rm H}(z)$, which sets the amplitude of the AP effect. To measure these parameters, a linear model for the link between the velocities of galaxies around voids and their density profile has been developed over the last decade \cite{Cai_2016,Hamaus_2016, Nadathur_2018,Nadathur_2018b, Hamaus2020:2007.07895, Hamaus_2022}.  Together with templates for the density and velocity dispersion profiles around voids and the AP and RSD parameters, this can be turned into a full model for the void-galaxy cross-correlation. 

The template method has been used to provide constraints for previous surveys, notably the Sloan Digital Sky Survey \cite{York2000:astro-ph/0006396} Baryon Oscillation Spectroscopic Survey (BOSS; \cite{Dawson_2013}) and extended BOSS (eBOSS; \cite{Dawson_2016}) data \cite[e.g.][]{Nadathur_2020, Hamaus2020:2007.07895,Woodfinden_2022,Woodfinden_2023}, the increased volume and precision of the DESI \cite{DESI2016a,DESI2016b} and Euclid surveys \cite{Laureijs_2011} means that we should revisit the method. Clearly, a template-based approach can be made more robust by allowing the shape of the distributions for which templates were adopted to vary with the model fitted. In other words the template method only has a limited range of validity around the models used to determine the templates. 

Improving the robustness of these void-galaxy cross-correlation function pipelines has been crucial. Reference \cite{Nadathur2019:1805.09349} introduced a method to remove the effect of RSD on void finding by performing reconstruction \cite{Eisenstein2007}, making RSD measurements more robust, as long as one trusts reconstruction on small scales. The AP effect on void finding was investigated in detail by \cite{Radinovic_inprep}, who found that it was important for the derived cosmological parameter errors. Other groups have considered varying assumed parameters in their pipelines, for example \cite{Hamaus2020:2007.07895} varied their choice of fiducial $\Omega_{\rm{m}}$, which would have the effect of including the AP effect on void finding. They found consistent results for different distance-redshift relationships. An emulator-based approach allows us to consistently include all of these effects.

Optimizations in $N$-body codes and increased computing power have enabled suites of high resolution and large volume cosmological simulations to be run and made available. Progress in machine learning techniques have facilitated the development of emulators: a way to forward model summary statistics given a small set of parameters \cite{Cranmer2020}. This technique primarily makes use of either Gaussian processes or neural networks, and has seen a wide variety of successes in the literature: ranging from the emulation of the matter power spectrum \cite{Heitmann2009:0902.0429, Angulo2021:2004.06245, Arico2021:2104.14568, Ramachandra_2021}, to two-point statistics of biased tracers \cite{Nishimichi2019:1811.09504, Kobayashi2022:2110.06969, Yuan_2022, Zhai2023:2203.08999, Lange2023:2301.08692, Chapman_2023, Hahn2023:2310.15246}, to higher-order and alternative clustering statistics \cite{storeyfisher2022aemulus, valogiannis2023precise, cuestalazaro2023sunbird, paillas2023cosmological, Massara2024:2404.04228, Hou2024:2401.15074}, notably with a wide range of cosmological and galaxy-halo connection models.

In this work, we extend the \textsc{sunbird} emulator \cite{cuestalazaro2023sunbird} to the problem of predicting the void-galaxy cross-correlation function. This allows us to fit the shape of the void-galaxy cross-correlation and velocity dispersion, removing the need for templates. We can also consistently include the AP and RSD effects on void finding, as void finding can be performed separately for each model in the training sample. We have trained our emulator across a wide range of cosmologies (85) and HOD models (100) from the \abacussummit\ suite of simulations \citep{Maksimova2021}. We then test the robustness of our emulator using cutsky mocks from the \nseries\ and \abacussummit\ suites. This is then applied to the CMASS galaxy sample of BOSS DR12, to directly obtain measurements of the standard $\Lambda$CDM cosmological parameters $\omega_{\rm cdm}$ \& $\sigma_8$. From these, we can constrain derived parameters including the linear growth rate of structure $f\sigma_8$, and the Alcock--Paczyński parameter $D_\mathrm{M}/D_\mathrm{H}$. We then compare these cosmological constraints with those obtained using the template method by \cite{Woodfinden_2022,Woodfinden_2023}.

Our paper is structured as follows: \cref{sec:data} outlines the data and mock catalogues used, \cref{sec:voidfinding} outlines the void finding and summary statistics we use. \Cref{sec:emulator} described our model training and the methodology used in the different stages of the \textsc{sunbird} pipeline, from training to covariance to emulation to inference. \Cref{sec:Results-mocks} summarizes our results from various tests run at different stages in the pipeline. The primary results are presented in \cref{sec:Results-data}. We discuss the results and their future impact in \cref{sec:discussion}. Finally, we state the key points of our paper in \cref{sec:conclusions}.

\section{Data and Mocks} \label{sec:data}

\subsection{Observational data}

We use the 12th data release of the Baryon Oscillation Spectroscopic Survey \citep[BOSS,][]{Dawson_2013}, which was part of the third stage of the larger Sloan Digital Sky Survey \citep[SDSS,][]{York2000:astro-ph/0006396}. We focus our analysis on the CMASS galaxy sample, which is comprised by luminous red galaxies that were selected based on SDSS multiband photometry \citep{Gunn1998:astro-ph/9809085, Gunn2006:astro-ph/0602326}. Although CMASS covers a redshift range $0.4 \lesssim z \lesssim 0.7$, we impose an additional cut $0.45 < z < 0.6$ to account for the fact that our simulation-based model is trained from simulations at a single redshift $z = 0.525$. For simplicity, we combine the measurements of the North and South Galactic Caps (NGC and SGC). 

In addition to the data catalogues, we use the random catalogs provided by the BOSS Collaboration, which follow the footprint and radial selection of the CMASS catalog, but with no intrinsic clustering. These randoms are used to estimate the overdensity field when identifying voids in the real data and calculating the clustering signal around them. We utilize random catalogues with 50 times the average number density of the CMASS galaxy catalog.

\subsection{Mocks}\label{subsection:Mocks}

We build the emulator using a series of mocks from the \abacussummit\ suite of $N$-body simulations \citep{Maksimova2021}. These mocks are used to generate the training data, to run the cosmological recovery tests and to evaluate the bias of the emulator. To evaluate the quality of our fits under survey-like conditions and weights, we use the 84 pseudo-independent \nseries\ mocks \citep{Gil-Marin_Nseries}, which match the volume, footprint, weights and radial number density of the BOSS survey, across the same redshift window we will use for CMASS ($0.45 < z < 0.60$) \citep{Reid2016:1509.06529}. We use the official suite of covariance mocks for the BOSS survey: The MultiDark simulation's series of 2048 \patchy\ mocks \citep{Kitaura_2014,Kitaura2016} that are used to compute the covariance matrices used for the \nseries\ mocks and for the CMASS data.  We will describe how each of these mocks are used in more detail below, and how they are used in different tests to our pipeline.

\subsubsection{\abacussummit\ mocks}\label{subsection:abacusmocks}

The \abacussummit\ simulations \citep{Maksimova2021} consist of a suite of $N$-body cosmological simulations run with the {\tt Abacus} code \citep{Garrison2021}, designed to meet the simulation requirements of the Dark Energy Spectroscopic Instrument \citep{DESI2016a,DESI2016b}. The suite is comprised of simulations of different sizes and resolutions, as well as multiple cosmological models. The {\it base} simulations follow the evolution of $6912^3$ dark matter particles in periodic boxes of $2 \Gpch$ on a side, with a mass resolution of $2 \times 10^9 \Msunh$. Dark matter halos are identified using {\tt CompaSO} algorithm \citep{Hadzhiyska2021}, which is a hybrid Friends-of-Friends and spherical overdensity algorithm.

In this work, we use 85 out of the 97 available cosmologies, which cover an extended $w_0w_{\rm a}$CDM parameter space, characterized by the physical baryon and cold dark matter densities ($\omegab$ and $\omegac$, respectively), the amplitude of matter fluctuations in spheres of $8\Mpch$ ($\sigma_8$), the spectral index and its running ($n_{\rm s}$ and $\alpha_{\rm s}$, respectively), the effective number of massless relics ($\Neff$), and the dark energy equation of state parameters $w_0$ and $w_{\rm a}$, which govern the parameterization $w(a) = w_0 + w_{\rm a}(1 - a)$, with $a$ the scale factor. In this model, $\Lambda$CDM is recovered when $w_0 = -1$ and $w_{\rm a} = 0$. The different simulations can be summarized as follows.
\begin{itemize}
    \item[-] {\tt c000}: Baseline \lcdm\ cosmology, with parameters that match the Planck 2018 best-fit model assuming base-\lcdm\ \citep{Planck2018Cosmo}.
    \item[-]{\tt c001}--{\tt c004}: Secondary cosmologies, including variations with a low $\omegac$, a thawing $w_0w_{\rm a}$ model, a high $\Neff$, and a low $\sigma_8$.
    \item[-]{\tt c013}: A base-\lcdm\ cosmology matching the Euclid Flagship 2 simulation \citep{Castander2024:2405.13495}.
    \item[-]{\tt c100--c126}: Linear derivative grid with single-parameter variations around the baseline.
    \item[-]{\tt c130--c181}: Emulator grid with a wider coverage of the parameter space.
\end{itemize}
All cosmologies assume a flat geometry, and the value of the dimensionless Hubble parameter $h$ is a derived parameter, calibrated to match the Cosmic Microwave Background (CMB) angular acoustic scale, $\theta_*$, measured by Planck 2018 \citep{Planck2018Cosmo}. In the next subsection, we will describe how we populate the dark matter haloes from \abacussummit\ by assuming a galaxy-halo connection semi-empirical model.

The \textit{small} simulations, used for the estimation of the covariance matrix, are run with the same resolution as the \textit{base} simulations, within periodic boxes that are $500\,\Mpch$ a side. These 1883 smaller boxes are only available for the baseline cosmology, {\tt c000}, with the same resolution as the \textit{base} catalogs, consisting of $1728^3$ particles. These mocks are also run for different realizations, i.e. different random seeds, to estimate the effect of cosmic variance. In \cref{subsec:hod_model}, we will describe how we populate the dark matter haloes from the \abacussummit\ simulations by assuming a galaxy-halo semi-empirical model.

There is a small offset between data and mocks as our CMASS sample has an effective redshift of $z_{\rm{eff}} = 0.525$, but the \abacussummit\ mocks have a redshift of $z=0.5$ ---  we do not expect this to be a significant concern, as we expect negligible redshift dependence on our summary statistics within this small redshift offset. Forthcoming \abacussummit\ lightcone mocks will solve this problem, and also the improve on the emulation by allowing for evolution within redshift bins rather than matching to a time-slice. 

\subsubsection{\nseries\ mocks}\label{subsection:Nseriesmocks}

The \nseries\ suite of simulations \cite{Gil-Marin_Nseries} consist of a series of 7 independent cubic periodic box mocks and 84 pseudo-independent cutsky mock catalogs designed to match the footprint, clustering and radial selection of the BOSS DR12 CMASS  galaxy sample. These cubic box mocks are run in a periodic box 2.6~$h^{-1}$Gpc a side with a mass resolution of $1.5 \times 10^{11} \ M_{\odot}$.  The halo catalogues are populated with an HOD model, with parameters chosen to best resemble the BOSS DR12 CMASS sample. The boxes are trimmed and rotated to produce 12 cutsky mocks that match the geometry of the CMASS sample, yielding 84 catalogues, and are run through the fiber assignment code used for BOSS to incorporate the effects of fiber collision \citep{Hanh2017}.  The \nseries\ mocks adopt a flat $\Lambda {\rm CDM}$ cosmology with $\Omega_{\rm m} = 0.286$, $\omegab = 0.0470$, $h = 0.70$, $\sigma_8 = 0.82$, and $n_{\rm s} = 0.96$.

\subsubsection{\patchy\ mocks}\label{sec:Patchymocks}
To construct the data covariance, we use the \textsc{big multidark-patchy} suite of mock catalogs, which we refer to as \patchy\ mocks \cite{Kitaura_2014,Kitaura2016,Rodriguez-Torres2016}. The approximated lightcone mocks are constructed through using Augmented Lagrangian Perturbation Theory (ALPT), which combines langrangian perturbation theory for larger scales and a spherical collapse halo model at smaller scales to obtain a halo catalog. These halos are then populated using a subhalo abundance matching (SHAM) code \patchy\ , which is calibrated to the match the 1,2, and 3-point clustering and bias of the BOSS LRG survey and the \textsc{big multidark} simulations. The mocks adopt the flat $\Lambda {\rm CDM}$ Planck2013 fiducial cosmology, given as: $\Omega_{\rm m } = 0.307115$, $\Omega_{\rm b} = 0.048$, $\sigma_8 = 0.8288$, $n_{\rm s} = 0.9611$, $h = 0.6777$.  These periodic boxes are then transformed into lightcone mocks as described by \cite{Kitaura2016}, and have been configured to match the angular footprint, halo occupation and redshift distribution of the BOSS DR12 survey. The lightcone mocks are then assembled by using the process described in \cite{Rodriguez-Torres2016}, wherein each of the redshift snapshots transformed to match the angular footprint and redshift distribution of BOSS DR12, with the same completeness and fiber collision weight codes applied. This is done to create mock lightcones of the LOWZ sample, the LOWZ+CMASS sample, and the CMASS sample, for both galactic caps (NGC, SGC). In this work, we use the CMASS sample of mocks. These 2048 mocks include both stellar mass completeness weights and fiber collision weights. These mocks are used to compute our data covariance, for both the \nseries\ cutsky void-galaxy cross-correlation function (VG-CCF), and for the CMASS void-galaxy cross-correlation function, across the redshift window of $0.45 < z < 0.60$.

\subsection{Galaxy-halo connection}
\label{subsec:hod_model}

We establish the link between dark matter haloes and galaxies using a halo occupation distribution (HOD) prescription. The HOD is a statistical model that assigns galaxies to dark matter halos based on properties of the host halos, most importantly the halo mass. Dependencies on other secondary halo properties can be included, such as concentration and local environment.

In this work, we adopt the base halo model from \cite{Zheng_2005}, in which the expected number of central galaxies on halos of mass $M$ can be written as
\begin{align}
    \langle N_{\rm c} \rangle(M) = \frac{1}{2} \left(1 + \mathrm{erf} \left(\frac{\log M - \log M_{\rm cut}}{\sqrt{2} \sigma} \right)  \right)\,,
\end{align}
where $\mathrm{erf}(x)$ is the error function, $M_{\rm cut}$ is the minimum mass required to host a central galaxy, and $\sigma$ is the slope of the transition from having zero to one central galaxy. The expected number of satellite galaxies is given by
\begin{align}
    \langle N_{\rm s} \rangle(M) = \langle N_{\rm c} \rangle(M) \left(\frac{M - \kappa M_{\rm cut}}{M_1} \right)^{\alpha}\,,
\end{align}
where $\kappa M_{\rm cut}$ corresponds to the minimum mass required to host a satellite galaxy, $M_1$ is the typical halo mass to host a single satellite, and $\alpha$ is the power law index for the number of satellites.

We generate 8,500 HOD parameter combinations in a Latin hypercube, and we assign 100 parameters to each of the \abacussummit\ cosmologies. We then populate the \abacussummit\ halos with galaxies using {\tt abacushod} \citep{Yuan2022b:2203.11963}, which is a highly efficient and modular {\tt python} code that implements multiple HOD prescriptions.

Whenever possible, our HOD catalogs are tuned to match the observed number density of galaxies from the CMASS sample at $0.45 < z < 0.6$, which is close to $3 \times 10^{-4}\, (\hMpc)^3$. When a certain combination of HOD parameters gives a number density of galaxies that is above our target, we randomly subsample the catalog to match the target, invoking an incompleteness fraction $f_{\rm ic}$. HODs with number densities below the target are kept as is, but they constitute a small fraction of the total number of HODs in our training sample.

The distant observer approximation is then applied to map the real-space positions of galaxies to their redshift-space positions, by using the peculiar velocity components to perturb the galaxy positions appropriately. This is done for each of the three lines of sight, giving three samples of the same mock with different redshift space distortions applied, which are then averaged to give the final measurement. This significantly reduces the noise in the quadrupole measurements, which are anti-correlated along the three lines of sight \citep{Smith2021:2007.11417}.

\section{Finding voids and measuring the void-galaxy cross-correlation} \label{sec:voidfinding}

The Alcock--Paczyński (AP) effect \cite{AlcockPaczynski1979} arises from the choice of fiducial cosmology being inconsistent with the true cosmology of the Universe. In this case, the distances across and along the observer's line of sight appear distorted, with a stretching that is governed to first order by the factors
\begin{equation} \label{eqn:AP}
    q_{\perp} = \frac{D_{\rm M}(z)}{D^{\rm fid}_{\rm M}(z)}, \ \ \  q_{\parallel}=\frac{D_{\rm H}(z)}{{D_{\rm H}^{\rm fid}(\rm{z})}} \,,
\end{equation}
where $D_{\rm M}(z)$ and $D_{\rm H}(z) = c/H(z)$ are the comoving angular diameter and Hubble distances to redshift $z$, respectively, and the superscript 'fid' denotes quantities in the fiducial cosmology. 

The AP effect is inherently present when one starts with a catalog of redshifts and assumes a wrong fiducial cosmology to convert them to distances, as is the case when working with real data. Since our goal is to train an emulator from simulations that can reproduce the clustering patterns of the real data, we manually distort the box coordinates using the factors from \cref{eqn:AP}, before void finding. More specifically, we rescale the box dimensions and galaxy coordinates across and along the line of sight by a factor $q_{\perp}$ and $q_{\parallel}$, respectively. These factors are determined for each mock and depend on the mock true cosmology and redshift, and our choice of fiducial cosmology, which is taken to be the {\tt c000} \abacussummit\ cosmology. A natural consequence of this stretching is that the simulation boxes are not necessarily cubic anymore, although they maintain periodicity across each face, with a period that is now given by the box length across that dimension times the corresponding AP stretching factor. Once the boxes are distorted, we only ever use the fiducial cosmology to perform subsequent actions, mimicking the process applied to actual data, where only a single fiducial cosmology is used.

We perform void finding using the {\tt voxel} algorithm implemented in the {\tt Revolver} code\footnote{\url{https://github.com/seshnadathur/Revolver}}. {\tt voxel} identifies underdense regions in simulations or real data without making assumptions about the shapes of voids, while still maintaining computational efficiency. The algorithm starts by painting galaxy positions to a rectangular grid\footnote{When applying the void finder to the CMASS data or the HOD mocks that have been stretched by the AP effect, this grid will generally be non-cubic. We have modified the void finder to handle non-cubic grid configurations, maintaining the periodic boundary conditions across each axis in the simulations case.} in comoving Cartesian coordinates, with a cell resolution $R_{\rm cell}$, which we set to $5 \Mpch$. This is converted to an overdensity grid, either by dividing by the average number of galaxies per cell in the periodic-box case, or by a grid that is painted with a random catalog that follows the survey geometry when working with observational data and the \nseries\ mocks. In the latter case, we mask those cells for which the number of random particles is below 75\% of the average number of randoms per cell. The overdensity field is then smoothed with a Gaussian filter of a characteristic length $\Sigmasm$, which we set to $10 \Mpch$ in this work. A watershed algorithm is then applied to classify the cells into different watershed regions. At this stage, each of those watershed regions is considered an individual void.

The center of a void is defined as the center of a voxel that has the lowest density among all voxels comprising the void. We also define an effective volume for each void, which is determined by the radius of a sphere that has the same volume as the sum of the volumes or all voxels or cells that compose the void. The final void catalog is then constructed by removing those voids with radii below $20 \Mpch$. We have chosen this cut to remove small, spurious voids that arise due to shot noise \cite{Neyrinck2008:0712.3049,Cousinou_2019}, and to ease comparison with previous studies in the literature \citep{Woodfinden_2022, Woodfinden_2023}.

We note that the void finder is applied directly to the redshift-space galaxy distribution. Previous works \citep{Chuang2017:1605.05352,Nadathur2019:1805.09349,Correa2022:2107.01314}, have shown that void finding in redshift space leads to additional selection effects on top of the standard galaxy RSD that are difficult to model analytically. \cite{Hamaus2020:2007.07895,Hamaus_2022} modify their analytic model with additional terms in order to try to absorb these effects. \cite{Correa2022:2107.01314} introduce correction factors to account for the bulk displacement of voids in redshift space, but also find that these do not account for the full effect of finding voids in redshift space. In contrast, \cite{Nadathur2019:1805.09349} apply a density-field reconstruction algorithm prior to void finding, which effectively removes RSD from the galaxy catalog. In our work, we use a neural-network emulator that automatically learns how the clustering identified in redshift space responds to changes in cosmology and HOD assumptions. Therefore, the complexities of void finding in redshift space are included in the emulator by construction. Thus we do not need to include extra factors to account for these effects, or run reconstructions, a major advantage of the emulator-based method.

Another advantage is that the void finder can be run on the simulations after the AP effect has been applied. This guarantees that any selection effects that AP might induce on the void-finding procedure are captured by the emulator, in the same way as it would appear in the real data if redshifts were converted to distances using only a fiducial cosmological model. This is motivated by the work of \cite{Radinovic_inprep}, who found that ignoring the AP effect at the void-finding level in the model can lead to biased estimates of the clustering signal around voids. To see how these could arise, suppose that the fiducial cosmological model were incorrect, such that the survey when analysed in terms of distances (rather than redshifts) is stretched along the line of sight, compared to across it. In this case, a void finder may preferentially find voids that are not intrinsically spherical, leading to a biased recovery of cosmological parameters.

\subsection{Void-galaxy cross-correlation}\label{subsec:vgccf}

Our main observable is the cross-correlation function between the void centers and the redshift-space galaxy positions, which we call the void-galaxy cross-correlation function (VG-CCF). This is calculated using {\tt pycorr}\footnote{\url{https://github.com/cosmodesi/pycorr}}, which is a wrapper around a modified version of the {\tt CorrFunc} pair-counting code \citep{corrfunc}. The VG-CCF is estimated using the natural estimator for the periodic simulation boxes where we can analytically calculate the expected density, and the Davis-Peebles estimator \citep{Davis&Peebles1983} for survey-like cutsky mocks and the CMASS data, 
\begin{eqnarray} \label{eq:estimators}
\xi_{\rm{DP}}(s) = \frac{\mathcal{VG}(s)}{\mathcal{VR}(s)} - 1 \,.
\end{eqnarray}
Here, $\mathcal{VG}(s)$ and $\mathcal{VR}(s)$ denote the normalized void-galaxy and void-random pair counts. We use the Davis-Peebles estimator rather than the Landy-Szalay estimator to avoid having to create an additional random catalogue with a distribution matching that of the voids. We note that the Davis-Peebles estimator is a formally unbiased estimator, and thus we expect this to add only a small amount of additional sensitivity to a mismatch between galaxy and random densities, as noted by the literature \cite{Pons_Borderia_1999,Vargas_Magana_2013}. To generate such a catalogue requires rerunning the void finder numerous times---a computationally expensive process. We bin the VG-CCF in $s$ and $\mu$, where $s$ is the scalar void-galaxy pair separation, and $\mu$ is the angle between the line of sight and the vector that connects the pair. We use 30 linearly-spaced bins in $s$ from $0.7 \Mpch$ to $120 \Mpch$, and 240 linearly-spaced bins in $\mu$ from $-1$ to $1$. We then decompose the correlation function into multipoles,
\begin{equation}
    \label{eq:multipoles}
    \xi_\ell (s) = \frac{2 \ell + 1}{2} \int_{-1}^1 \rm d \mu \, \xi(s, \mu) P_\ell (\mu),
\end{equation}
where $P_\ell (\mu)$ is the Legendre polynomial of order $\ell$. In particular, we focus on the monopole $\ell = 0$ and quadrupole $\ell = 2$ moments, as the low signal-to-noise ratio of the hexadecapole remains prohibitive for a clustering analysis.

We calculate the VG-CCF for each of the HOD catalogs described in the previous section. When doing so, we apply the RSD along the three simulation axes and measure the VG-CCF for each of them separately, averaging the clustering measurements to increase the signal-to-noise ratio, as in \citep{Smith2021:2007.11417}. We additionally measure it for the \nseries\ and \patchy\ mocks, as well as the CMASS data.

\subsection{Data covariance} \label{sec:Cdata}

We computed the variance associated with the data vector comprising sample variance and shot noise by taking the 2048 realizations of the \textsc{big multidark-patchy} lightcone mocks, and computed the monopole and quadrupole of each realization, across a redshift window of $0.45 < z < 0.60$. As previously discussed in \cref{sec:Patchymocks}, these mocks are made to match the clustering, angular footprint and number density of the BOSS DR12 samples, and we used the CMASS NGC sample. We computed the redshift-space VG-CCF with Davis-Peebles estimator and used their random catalog with 50 times the galaxy number density. We follow the binning of void-galaxy separation of \cite{Woodfinden_2022}: 30 bins for $0.7\leq s \leq 120 \rm \ \Mpch$. The resulting covariance $\tensor{C}^{\rm data}$ is displayed in the bottom panel of \cref{fig:Covariance}.

\section{Building an emulator}  \label{sec:emulator}

We build an emulator of the VG-CCF using the HOD catalogues described in \cref{subsec:hod_model}. We train a fully-connected neural network to learn the dependence of the VG-CCF on the cosmological and HOD parameters using the \textsc{sunbird}\footnote{\url{https://github.com/florpi/sunbird}} package \citep{cuestalazaro2023sunbird}. We begin by splitting the HOD catalogues into a training, validation, and test set. We reserve the {\tt c000}-{\tt c004} and {\tt c013} cosmologies described in \cref{subsection:abacusmocks} for the test set, as they correspond to locations in the parameter space where we might expect the emulator to fail most strongly. The test sample does not need to be representative, and this should therefore be considered conservative. From the remaining cosmologies, we randomly split the simulations, leaving 74 of them for training, and 5 for validation. The inputs of the neural network are the cosmology and HOD model parameters, normalized to be within 0 and 1, and the output are the concatenated monopole and quadrupole moments of the VG-CCF, also normalized. We use Sigmoid Linear Units (SiLU) as the activation function and adopt a mean squared error as the loss function, 
\begin{equation}   \label{eq:MSE} 
   {\rm MSE} =  \frac{1}{N} \sum_{\rm{i}=1}^{N}(\rm{X_i}^{\rm{theo}}(\mathcal{C},\mathcal{G})-\rm{X_i}^{\rm{obs}})^2 \,,
\end{equation}
where $\rm{X_i}^{\rm{obs}}$ is the observed data vector, $\rm{X_i}^{\rm{theo}}(\mathcal{C},\mathcal{G})$ is the theoretical prediction, which depends on the cosmological ($\mathcal{C}$) and HOD ($\mathcal{G}$) parameters, and $N$ is the batch size, which we set to 256. We optimise the neural network hyperparameters using {\tt optuna}\footnote{\url{https://github.com/optuna/optuna}}, varying the learning rate, number of layers, number of hidden units per layer, dropout rate, and the weight decay, and selecting the model that minimizes the validation loss. These hyperparameters are used in regularization routines that help improve a neural network's generalization ability by reducing overfitting.

\begin{table}
    \centering
    \small
    \begin{tabular}{ |l|l|l| }
    \hline
        Parameter & Prior & Interpretation \\
        \hline
         $\omega_{\rm cdm}$ & $\mathcal{U} (0.1032,0.140)$ & Physical cold dark matter density \\
         $\omega_{\rm b}$ & $\mathcal{U} (0.0207,0.0243)$ & Physical baryon density\\
         $n_{\rm s}$ & $\mathcal{U} (0.9012,1.025)$ & Tilt of the primordial power spectrum\\
         $\sigma_8$ & $\mathcal{U} (0.678,0.938)$ & r.m.s of matter fluctuations in spheres of $8 \Mpch$\\
        $w_0^*$ & $\mathcal{U} (-1.22,-0.726)$ & Present-day dark energy equation of state parameter\\
        $w_{\rm a}^*$ & $\mathcal{U} (-0.628,0.621)$ & Redshift evolution of the dark energy equation of state\\
         \hline
         $\log M_{\rm cut}$ & $\mathcal{U} (12.4,13.3)$ & Minimum halo mass to host a central galaxy \\
         $\log M_1$ & $\mathcal{U} (13.2,14.4)$ & Typical halo mass to host one satellite galaxy\\
         $\alpha$ & $\mathcal{U} (0.7, 1.5)$ & Power-law index of the expended number of satellites\\
         $\kappa$ & $\mathcal{U} (0.0, 1.5)$ & Regulation of the minimum halo mass to host a satellite \\
         $\log \sigma$ & $\mathcal{U} (-3.0, 0.0)$ & Slope of the transition from zero to one central galaxy \\
         \hline
    \end{tabular}
    \caption{Prior distributions used during the parameter inference stage. Parameters marked with a star are fixed to their default base-$\Lambda$CDM values when not considering these model extensions.}
    \label{tab:priors}
\end{table}

\subsection{Covariance associated with the emulator} \label{sec:covariance}

In order to estimate the covariance between data and emulator-based model, we follow \cite{cuestalazaro2023sunbird} and split into three components that arise from: the ability of the emulator to reproduce the same-realization simulations used for training $\tensor{C}^{\rm emu}$, the difference between the single-realization simulations and the true model $\tensor{C}^{\rm sim}$, and the effects of cosmic variance and shot noise in the data $\tensor{C}^{\rm data}$. The way that we calculate $\tensor{C}^{\rm data}$ was described in \cref{sec:Cdata}.

\subsubsection{Simulation covariance}

\begin{figure}[!h]
    \centering
    \addtolength{\leftskip}{-5cm}
    \addtolength{\rightskip}{-5cm}
    \includegraphics[width=1.\textwidth,keepaspectratio]{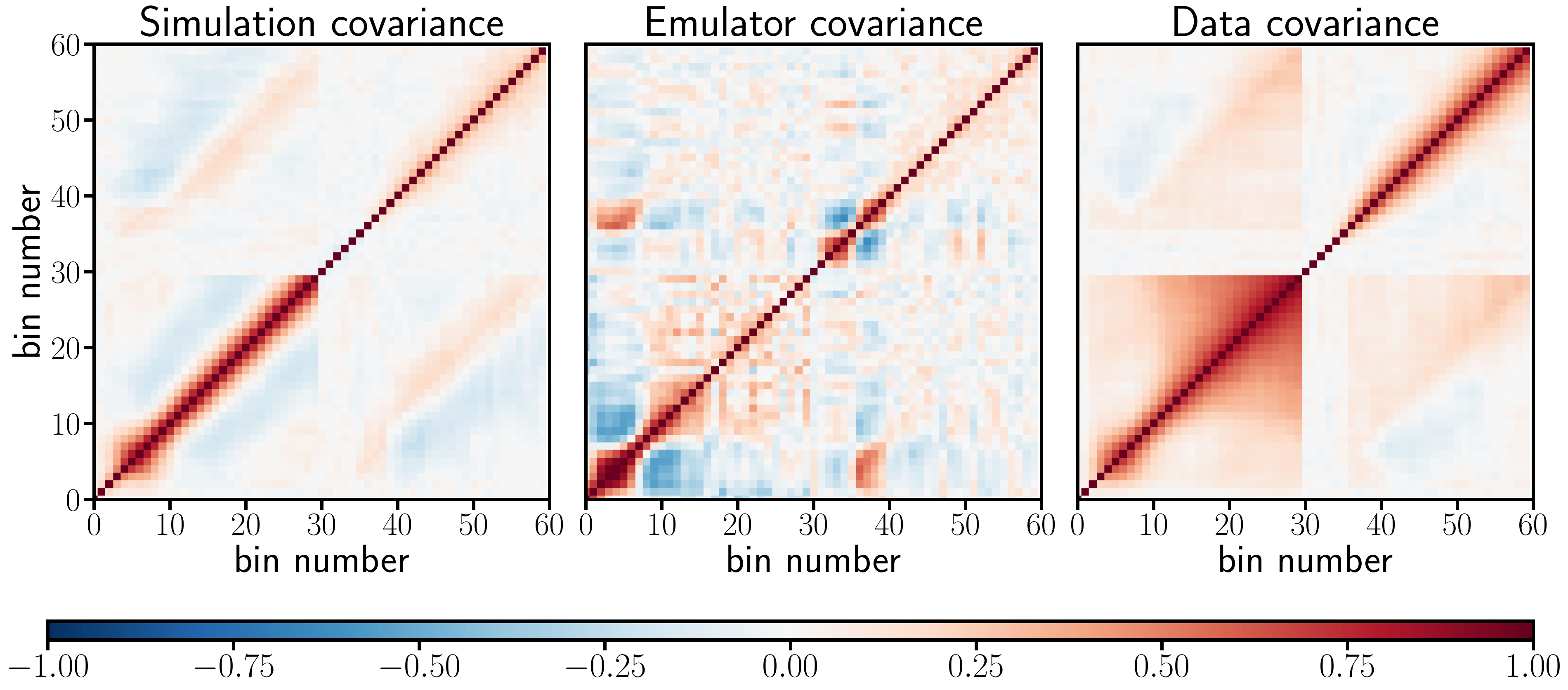}
    \caption{Correlation matrices associated to the different components that contribute to the total error budget in our likelihood analysis. The simulation covariance (left) accounts for cosmic variance due to the finite volume of the \abacussummit\ simulations used to train our model. The emulator covariance (center) quantifies the systematic error in our model due to an imperfect emulation of the multipoles of the void-galaxy cross-correlation function. The data covariance (right) was computed from the monopoles and quadrupoles of the void-galaxy cross-correlation function, obtained from 2048 \patchy\ mocks  $0.45 < z < 0.60$. }
    \label{fig:Covariance}
\end{figure}

Since our emulator is trained from \abacussummit\ \textit{base} boxes with a fixed phase realization, its predictions do not factor in the contributions from cosmic variance due to the finite simulation volume. To quantify this error contribution, we use the \abacussummit\ \textit{small} boxes, which were run with the baseline \abacussummit\ cosmology and include 1883 different phase realizations. We calculate the VG-CCF multipoles from each realization using a fixed HOD model, with parameters calibrated to minimize the $\chi^2$ between the predicted multipoles and those measured from the BOSS CMASS sample. The procedure for measuring the CCFs follow the description from \cref{subsec:vgccf}, averaging along each of the three lines of sight for each box. The covariance is then calculated using the standard method
\begin{equation}\label{eq:Csim}
    \tensor{C}^{\rm sim} = \frac{1}{(N_{\rm s} -1)} \sum_{\rm{i}=1}^{N_{\rm s}} (\overline{\vect{{d}}} - \vect{d_{\rm i}})(\vect{\overline{{d}}} - \vect{d_{\rm i}})^{T}
\end{equation}

Where $\vect{\overline{d}}$ is the data vector (the concatenated monopole and quadrupole) averaged across all simulations ($N_{\rm s}$), and $\vect{d_{\rm i}}$ is the data vector for the $i$th simulation. These terms in \cref{eq:Csim} are used to estimate the simulation covariance, shown in the left panel of \cref{fig:Covariance}. Before appending this component into the overall covariance matrix, we rescale the covariance to match the volume of an \abacussummit\ \textit{base} box, dividing the term by a factor of 64. While such a volume scaling is approximately correct, it neglects some nuances such as the super-sample covariance, because both smaller boxes and larger boxes have mean overdensity set to exactly the expected background level, rather than allowing for fluctuations in this \cite{Howlett_2017}. At the level of our analyses, we do not expect these issues to be important.

\subsubsection{Emulator covariance} \label{sec:cov_emu}

\begin{figure}
    \centering
    \begin{tabular}{cc}
      \includegraphics[width=0.44\textwidth]{AbacusResidual_c000_new.pdf}   &  \includegraphics[width=0.44\textwidth]{AbacusResidual_c001_new.pdf}\\
      \includegraphics[width=0.44\textwidth]{AbacusResidual_c002_new.pdf}   &  \includegraphics[width=0.44\textwidth]{AbacusResidual_c003_new.pdf}\\
      \includegraphics[width=0.44\textwidth]{AbacusResidual_c004_new.pdf}   &  \includegraphics[width=0.44\textwidth]{AbacusResidual_c013_new.pdf}
    \end{tabular}
    \caption{Emulator predictions of the monopole and quadrupole, for each of our six test cosmologies and averaged across 100 HODs, with their residuals in the panel below. We compute the covariance from the volume-adjusted \abacussummit\ 500~$h^{-1}$Mpc covariance box, scaled to match a single measurement of the monopole and quadrupole from a 2~$h^{-1}$Gpc box. Because we stack voids this will conflate the interior of larger voids with the edges of smaller voids, and it is hard to judge the behaviour of individual voids of different sizes from this plot. }
    \label{fig:AbacusResiduals}
\end{figure}

\begin{figure}[htpb]
    \centering
    \includegraphics[width=\textwidth]{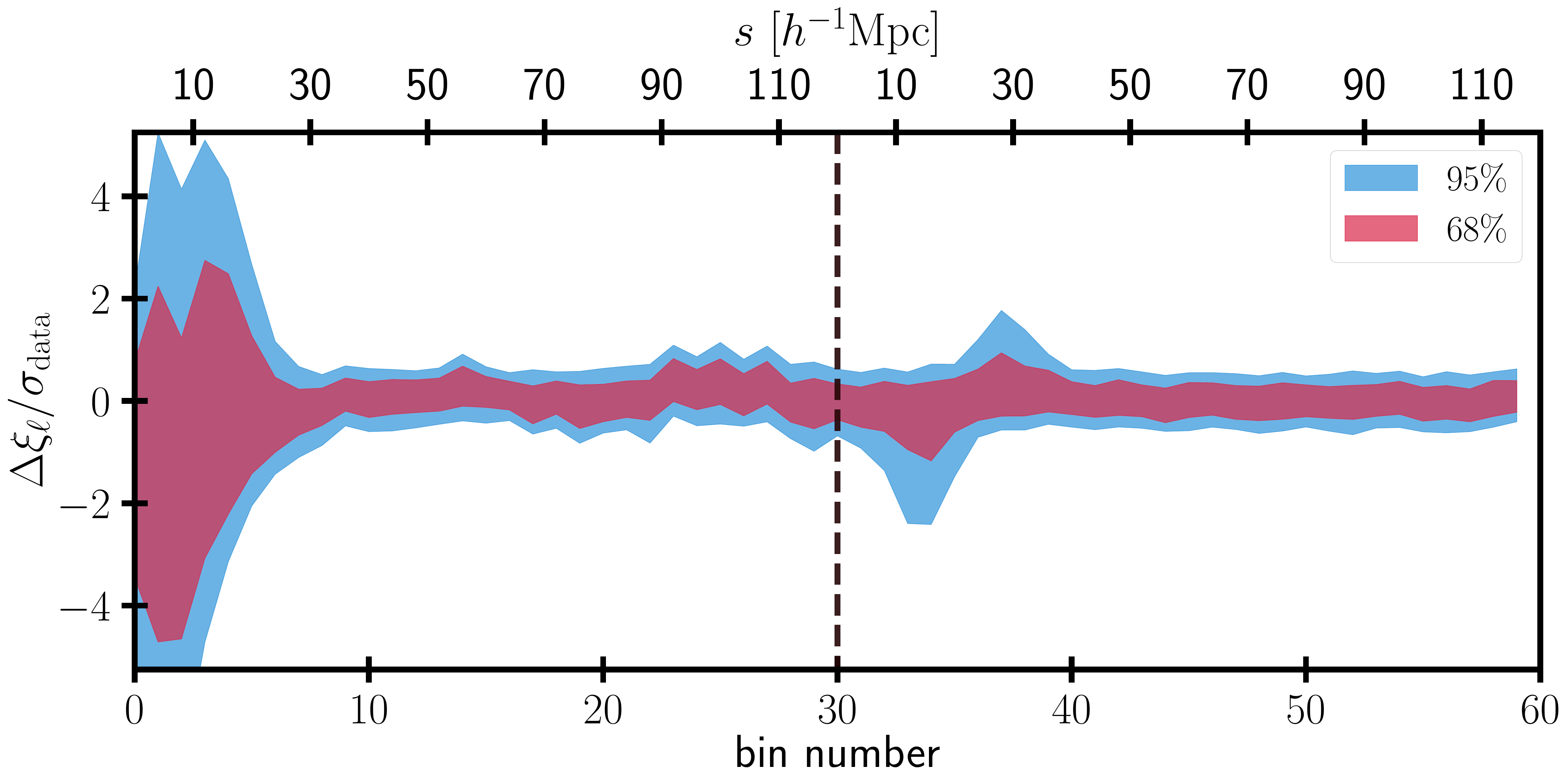}
    \caption{68 and 95 percentiles (red and blue, respectively) associated to the difference between the emulator predictions and the test data for the VG-CCF monopole (bin numbers from 0 to 30) and quadrupole (30 to 60), measured across 600 test simulations. This is shown in units of the error bars from the CMASS data, calculated from the 
    \patchy\ covariance mocks. The bin numbers associated to each multipole cover a scale range $0.7\leq s \leq 120 \rm \ \Mpch$, shown on the upper x-axis. We note that the performance of the emulator is worse for bins below $25 \Mpch$ for the monopole and $30 \Mpch$ for the quadrupole. We will discuss the emulator error in subsequent sections and its importance to future emulator work.}
    
    \label{fig:AbacusBiasTest}  
\end{figure}

To study the systematic error in our model due to an imperfect emulation, we examine six test cosmologies with 100 HODs each, giving us a sample of 600 mocks in total. The test cosmologies include {\tt c000}, a baseline $\Lambda \rm{CDM}$ model using the Planck 2018 parameters; {\tt c001}, a $\Lambda \rm{CDM}$ model using the WMAP9+ACT+SPT parameters from \cite{calabrese2017}; {\tt c002}, a $w\rm{CDM}$ model with thawing parameters; {\tt c003} a $\Lambda \rm{CDM}$ model with a higher $\Neff$; {\tt c004} a baseline $\Lambda {\rm CDM}$ model with low $\sigma_8$; and {\tt c013}, a simulation run with the Euclid Flagship 2 $\Lambda \rm{CDM}$ cosmology. For each of these mocks, we make model predictions using the true parameters of the simulations, and compare them against the measured multipoles from the simulations. \cref{fig:AbacusResiduals} shows how well the model is able to reproduce the data, where we have averaged the measurements for each cosmology across all 100 HOD models. We note the presence of small scale differences between the data and the model exceeding 3$\sigma$ for a few bins, particularly in {\tt c002, c003, c013}. These variations will be accounted for in the emulator covariance, which we discuss below.

In \cref{fig:AbacusBiasTest}, we show the 68 and 95 percentiles for the difference between the model prediction and the test data across all mocks. To assess how this systematic model uncertainty compares to the errors from the real observations, the vertical axis is shown in units of the standard deviation of the CMASS multipoles, as measured from the \patchy\ covariance mocks. Our emulator achieves high accuracy at most scales, with the emulation error being smaller than the error from the data across a wide scale range. The error budget is dominated by the systematic emulator uncertainty at small scales ($s \leq 20 \Mpch$), limiting the improvement from the increased data precision in this regime. The fractional accuracy of the emulator predictions remains similar across the full scale range, and therefore, it becomes dominant with respect to the data error as we approach smaller separations. Future work might be able to improve on this aspect by applying non-linear transformations to the input training data or weigh scales differently during the training process.

To quantify the emulator's contribution to the total covariance, we compute the covariance matrix associated to the difference between the model predictions and the test measurements, $\vect{d}^{\rm{meas-pred}} = \vect{d}^{\rm{meas}} - \vect{d}^{\rm{pred}}$, concatenating the monopole and quadrupole into a single data vector:
  
\begin{equation}\label{eq:EmuCov}
    \tensor{C}^{\rm emu} = \sum^{\rm n_{test}}_{\rm i=1} \left( \vect{{d}}^{\rm meas-pred}_{i}-\overline{\boldsymbol{d}^{\rm meas-pred}_{i}}\right) \left( \vect{d}^{\rm meas-pred}_{i}-\overline{\boldsymbol{d}^{\rm meas-pred}_{i}}\right)^{T}\,.
\end{equation}

Here, an overline denotes the average data vector across all 600 mocks. The resulting covariance matrix is shown in the middle panel in \cref{fig:Covariance}. While this matches the procedure followed in \cite{cuestalazaro2023sunbird}, this estimation of the emulator error can be deemed conservative. The variance of these residuals calculated across the full set of test simulations is sensitive to outliers in the distribution, which can appear due to the emulator being less accurate at extreme regions of the parameter space that are not densely covered by the training data. This can potentially overestimate the true emulator uncertainty in regions of the parameter space that are well sampled by our simulations.

\subsection{Inference with the emulator and MCMC} \label{sec:mcmc}

We use the Markov Chain Monte-Carlo (MCMC) technique to explore the posterior surface for the parameters of interest. We use the {\tt numpyro}\footnote{\url{https://github.com/pyro-ppl/numpyro}} implementation of the adaptive Hamiltonian Monte Carlo (HMC) sampler. We cross-referenced the posteriors obtained with HMC with other samplers, and found strong agreement.

For the cosmology and HOD parameters, we kept our priors uniform, as specified by \cref{tab:priors}. For the cosmological parameters, these match the bounds defined by \abacussummit\ for their simulations. We evaluated the likelihood 
\begin{equation} \label{eq:Likelihood}
    \log \mathcal{L} = \left(\rm \vect{X}^{\rm obs} - \rm \vect{X}^{\rm theo}(\mathcal{C},\mathcal{G})\right) \tensor{C}^{-1} \left(\rm \vect{X}^{\rm obs} - \rm \vect{X}^{\rm theo}(\mathcal{C},\mathcal{G})\right)^{\rm T}\,,
\end{equation}
where $\rm{X^{obs}}$ is the measured data vector. $\rm{X}^{\rm theo}(\mathcal{C},\mathcal{G})$ is the model predicted by the emulator for cosmological parameters $\mathcal{C}$ and HOD parameters $\mathcal{G}$. The covariance $\rm{C}$ has been adjusted by applying the correction proposed by \cite{Percival_2022} to each component of the covariance according to how it was calculated.

\subsection{Angular CMB scale prior} \label{sec:theta*}

Each \abacussummit\ simulation used to train our emulator adopted a value of dimensionless Hubble parameter $h$ that gave the same angular CMB peak position $\theta_{*}$, matching the best-fit value ($100 \theta_{*} = 1.0435133$) obtained by Planck 2018 \cite{Planck2018Cosmo}. We impose a prior on $\theta_{*}$ to ensure that our emulator is used  within the same parameter space as it was trained. Importantly, using \abacussummit, we can only emulate cosmologies that match this value of $\theta_{*}$.

Consequently, we do not have $h$ as a free parameter in our fits, but determine it at each chain point from the other parameters and the fixed value of $\theta_{*}$. 

When we compare the emulator's posteriors with those obtained from the template method of \cite{Woodfinden_2022,Woodfinden_2023}, we must apply the same prior on $\theta_{*}$ to the template's MCMC chains. In both the case of the emulator and the template method, the inclusion of CMB information through $\theta_{*}$ is a significant source of constraining power. This means that our constraints are not exclusively sourced from late-time void-galaxy clustering, but also from the CMB. No CMB information other than the constraint on $\theta_{*}$ is included in this analysis, and we note that the prior on $\theta_{*}$ is well-motivated, as it is well constrained by Planck to within $0.03\%$ precision, and is robust under different cosmological models \cite{Planck2018Cosmo}.

\begin{table}[tbp]
    \centering
    \resizebox{\columnwidth}{!}{%
    \begin{tabular}{|c|cccccccc|}
    \hline
       sim  & $\omega_{\rm b}$  & $\omega_{\rm cdm}$ & $N_{\rm ur}$ & $n_{\rm s}$ & $\alpha_{\rm s}$ & $\sigma_8$ & $w_0$  & $w_{\rm a}$\\
       \hline
       c000  & 0.02237 & 0.1200 & 2.0328 & 0.9649 & 0.0 & 0.807952 & -1.0 & 0.0 \\
       c001  & \textbf{0.02242} & \textbf{0.1134} & 2.0328 & \textbf{0.9638} & 0.0 & \textbf{0.776779} & -1.0 & 0.0 \\
       c002  & 0.02237 & 0.1200 & 2.0328 & 0.9649 & 0.0 & \textbf{0.808189} & \textbf{-0.7} & \textbf{-0.5} \\
       c003  & \textbf{0.02260} & \textbf{0.1291} & \textbf{2.6868} & \textbf{0.9876} & 0.0 & \textbf{0.855190} & -1.0 & 0.0 \\
       c004  & 0.02237 & 0.1200 & 2.0328 & 0.9649 & 0.0 & \textbf{0.749999} & -1.0 & 0.0 \\
       c013  & \textbf{0.02200} & \textbf{0.1206} & 2.0328 & \textbf{0.9600} & 0.0 & \textbf{0.813715} & -1.0 & 0.0\\
       \hline
    \end{tabular}%
    }
    \caption{Table of cosmologies used in the test set. Bold values highlight changes from the fiducial {\tt c000} case. The full description of these simulations, the cosmologies used and the way these simulations are obtained can be found in \cite{Garrison2021}.}
    \label{tab:testcosmos}
\end{table}

\section{Testing the pipeline using mock catalogues}\label{sec:Results-mocks}

We train the emulator as specified in \cref{sec:emulator}, and quantify the emulator's ability to match the concatenated monopole and quadrupole for the \abacussummit\ test cosmologies through $\tensor{C}^{\rm emu}$, as described in \cref{sec:cov_emu}.

We now consider the ability of our full pipeline to recover the cosmological parameters of interest. We start by considering the \abacussummit\ simulations used to test the emulator in \cref{sec:cov_emu}, and then extend this with fits to the \nseries\ cutsky mocks, which include BOSS fiber collision effects and geometry --- properties which were not part of our training set. Finally, once we have shown that the cosmological recovery stages of the pipeline are robust, we apply the same pipeline to the BOSS CMASS DR12 NGC sample (see \cref{sec:Results-data}). 

\subsection{\abacussummit\ based tests}
 \label{sec:abacus-tests}

For each \abacussummit\ mock considered, we infer the posterior of the cosmological and HOD parameters; where the mock data vector is compared to those generated by the emulator.
We consider the same six test cosmologies {\tt c000, c001, c002, c003, c004, c013} used to construct $\tensor{C}^{\rm emu}$ in \cref{sec:cov_emu}. For each, we select HOD parameters that best fit the BOSS CMASS sample. We utilize the same binning as the CMASS sample, with 30 separation bins for the monopole and quadrupole, for $0.7 \Mpch \leq s \leq 120 \Mpch$. We run our inference with the priors prescribed by \cref{tab:priors}.

\begin{figure}
    \centering
    \includegraphics[width=\columnwidth]{c000_fullpost_new.pdf}
    \caption{Full posterior of the {\tt c000} test cosmology. True values for the parameters are shown by the solid black lines. Overall, we find that the posterior range for the HOD parameters $\log M_{\rm cut}$, $\log M_1$, $\log\sigma$, $\alpha$, $\kappa$ consistently hit the limits of our priors and are uncorrelated with our cosmological parameters. The full posteriors for the other test cosmologies can be found in \cref{sec:Appendix}. We then marginalize over these HOD parameters for the remainder of our analysis. }
    \label{fig:c000_full_post}
\end{figure}

\begin{figure}
    \centering    
    \hspace*{-2cm}
    \includegraphics[width=0.7\textwidth]{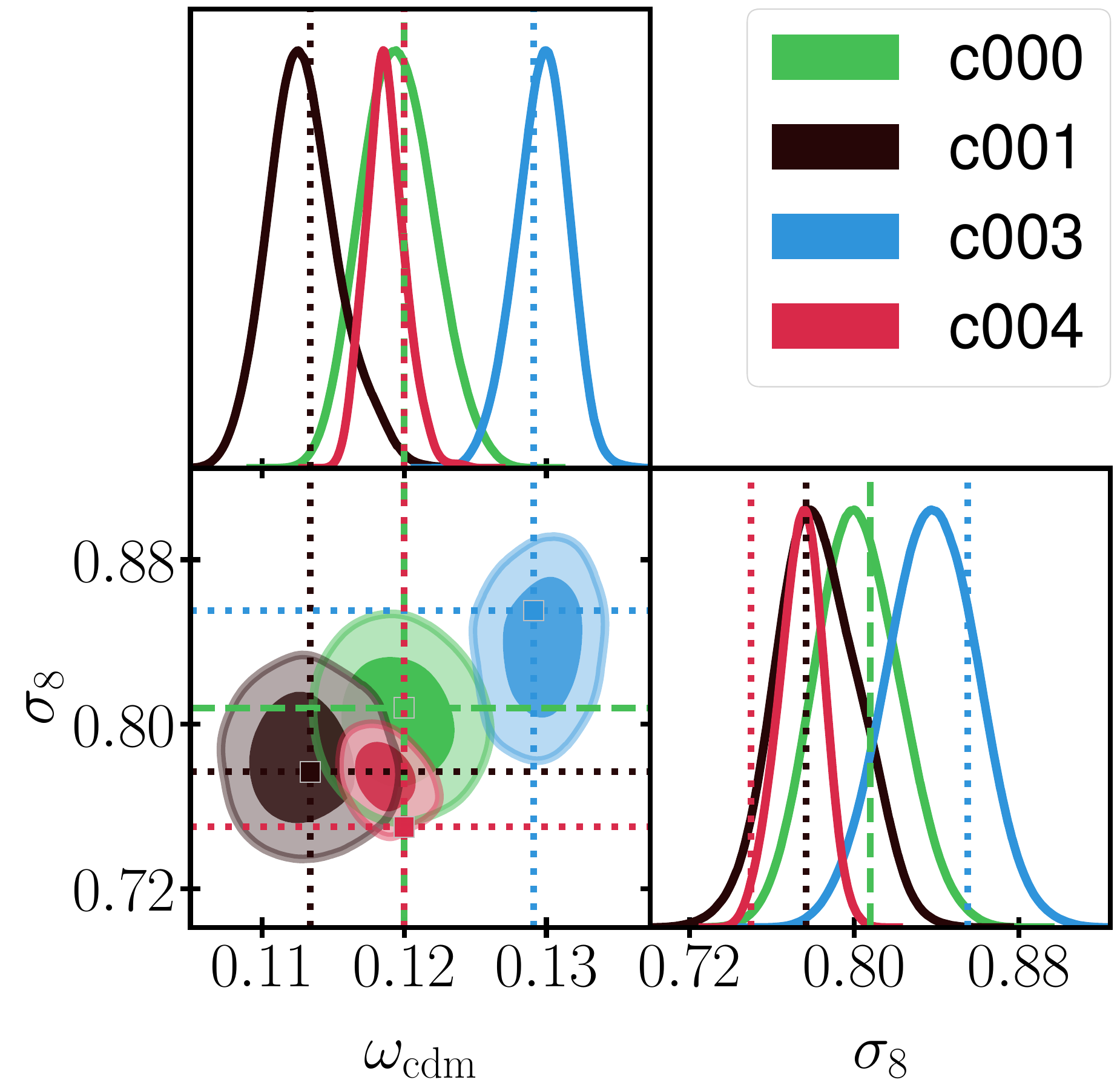}
    \caption{Posteriors obtained from the emulator, marginalized over $\omega_{\rm b}$, $n_{\rm s}$ and the HOD parameters, using the priors described in \cref{tab:priors}, with the test cosmologies and their true values for $\omega_{\rm cdm}$, $\sigma_8$ from \cref{tab:testcosmos}. This figure confirms that our emulator can successfully recover the true cosmological parameters after emulating the VG-CCF of our test set. Notably, these test cosmologies were chosen locations in the parameter space where we expected the emulator to fail most strongly.}
    \label{fig:AbacusTestPosterior}
\end{figure}

We first fit {\tt c000}, the baseline cosmology of \abacussummit. The full posterior distribution is shown in \cref{fig:c000_full_post}. \cref{sec:Appendix} contains the full posteriors of all of the test cosmologies.  We find that the sensitivity of the emulator to HOD parameters is limited---all five parameters hit the limits of our priors. We conclude that the VG-CCF is uninformative in terms of constraining the galaxy-halo connection, and we marginalize over these parameters for the remainder of this analysis. Our model does obtain constraints for $n_{\rm s}$, but since this quantity is highly constrained by the CMB and we do not expect much sensitivity from the VG-CCF to it, most of the constraints we present throughout the rest of the paper are marginalized over it. Our analysis will focus on obtaining cosmological parameters and derived quantities that pertain to the large scale structure evolution at late times. We find that for all six of our test cosmologies unbiased constraints can be obtained for $\omegac$ and $\sigma_8$. This is underscored by \cref{fig:AbacusTestPosterior}, where marginalized constraints for $\omegac$ and $\sigma_8$ are shown for four test cosmologies that vary these parameters, with the marginalized posteriors all falling within $2\sigma$ of the true parameter values. This figure explicitly shows that despite adopting a fiducial cosmology that does not resemble the one used in the box (e.g. {\tt c003}) we can successfully recover the correct cosmological parameter values for $\omegac, \sigma_8$.

\subsection{\nseries\ based tests} 
 \label{sec:nseries-tests}

\begin{figure}
    \centering
    \includegraphics[width=\columnwidth]{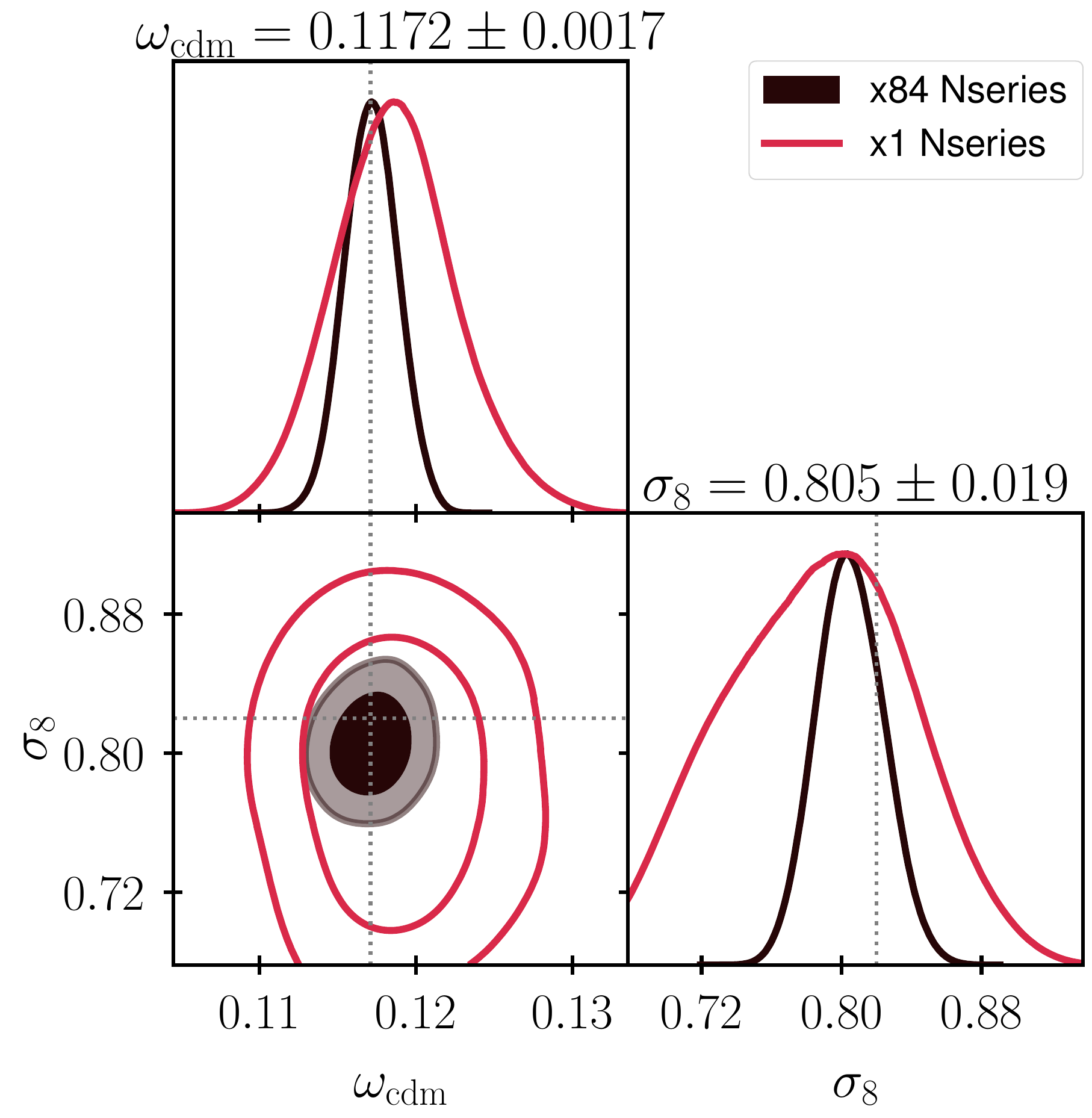} 
    \caption{Marginalized posteriors of the emulator's predictions of the \nseries\ cosmology, with the errors scaled to the BOSS volume (black) and 84 times the BOSS volume (red), with the best fit for the x84 chain annotated above. We find that in both cases the emulator can successfully predict $\omegac$ and $\sigma_8$ to within 1 $\sigma$. This confirms that the emulator's predictions are robust---and it can successfully recover cosmological parameters even when survey holes, geometry and fiber collision weights were not included in the training data.}
    \label{fig:Nseries_posterior}
\end{figure}

\begin{figure}
    \centering
    \includegraphics[width=\columnwidth]{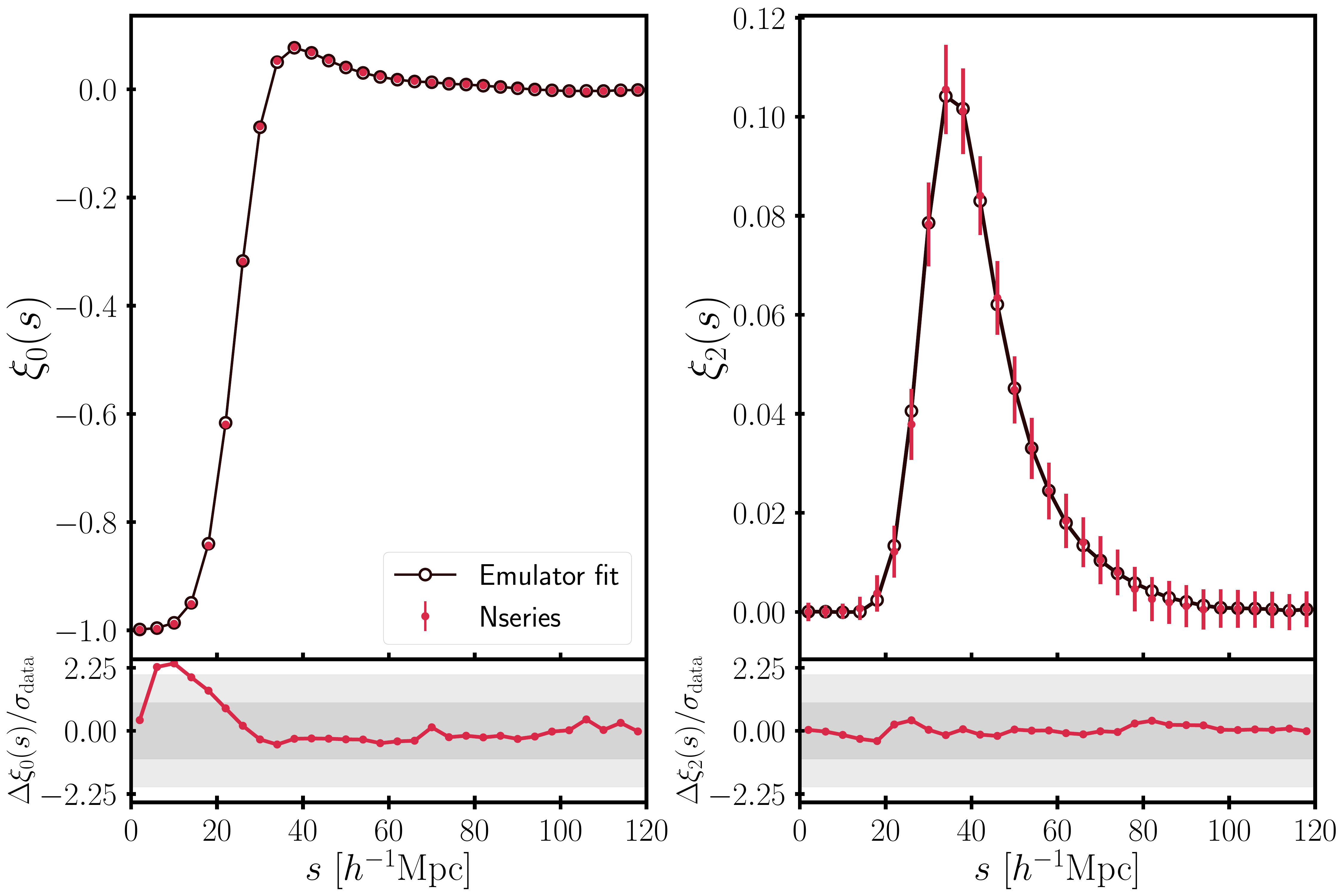}
    \caption{Predicted monopole and quadrupole from the \nseries\ cutsky mocks, with an x84 volume scaling applied to limit cosmic variance. Overall, we find that even with some variation in the recovered cosmological parameters, where we obtain $\Omega_{\rm{m}}$ to $1.3\sigma$ of the \nseries\ value and $\sigma_8$ to within $0.8 \sigma$ of the \nseries\ value, the monopole and quadrupole from the emulator's prediction closely match the monopole and quadrupole obtained from running the void finder on the 84 \nseries\ cutsky mocks.}
    \label{fig:Nseries_predictions_res}
\end{figure}

The recovery tests we run using the \nseries\ mocks are similar to those described in the previous section, with a few key differences: i) there is a single \nseries\ test cosmology and HOD model we infer ii) the \nseries\ mocks are run with a different $N$-body code and galaxy-halo connection model than one used in our training data and iii) the \nseries\ mocks are cutsky mocks, not cubic boxes iv) the cutsky mocks include approximate fiber collision and completeness weights \cite{Reid2016:1509.06529} and v) the radial number density is not uniform. Points iii)-v) are particularly salient, as they help us test the robustness of the pipeline with data that more closely resembles spectroscopic survey data. The \nseries\ mocks are not full light-cones, and so the growth is stepwise corresponding to the boxes from which the mocks were created. The geometry does match that of the BOSS sample, and the AP effect smoothly varies across the mocks, rather than being fixed at an effective value. Thus we consider that fitting to these offers a rigorous test of the method.

We fit the mean of the \nseries\ data vector, and consider errors scaled for either the total volume (labelled as x84 in \cref{fig:Nseries_posterior}) or the volume of a single mock (labeled as x1 in \cref{fig:Nseries_posterior}). The same priors are used as in the \abacussummit\ inference, and we obtain the 2D posterior distribution, marginalizing over all parameters except $\omegac$ and $\sigma_8$. We can then obtain $h$,  $\Omega_{\rm{m}}$ and compute $D_\mathrm{M}/D_\mathrm{H}$ and $f\sigma_8$, along the chain, and compare these to the values obtained from the true \nseries\ cosmology. The 2D posterior is shown in \cref{fig:Nseries_posterior}. We find that even when rescaling by the total volume (84 times the \nseries\ mock volume), the true values lie within the $2\sigma$ marginalised credible intervals. For the cosmological parameters, we obtain $\Omega_{\rm{m}}$ to $1.3\sigma$ of the \nseries\ value and $\sigma_8$ to within $0.8 \sigma$ of the \nseries\ value.

The residuals between the emulator's maximum likelihood solution and the mock multipoles are shown in \cref{fig:Nseries_predictions_res}. The residuals are shown in units of the standard deviation of the data vector, derived from the \patchy\ mocks. We find the residuals for the monopole and quadrupole lie within $2\sigma$, with the exception of two points, both in the monopole at scales below $20 \Mpch$. The bias seen at smaller scales is similar to the one seen in the tests with the \abacussummit\ data. Taken together with the results of the cosmological recovery, this indicates that our emulator is robust to changes in survey geometry, weights and $N$-body codes, with the caveat that the model might slightly mismatch the clustering amplitude at very small scales.

\section{CMASS data fits}
 \label{sec:Results-data}
 
Having tested the emulator on two sets of mocks, and shown that our pipeline is robust and accurate, we now fit to the BOSS CMASS DR12 galaxy catalog.

For template based models, fits are typically made to the multipoles using a compressed set of parameters quantifying the AP and RSD effects \cite[e.g.][]{Woodfinden_2022}. For our emulator-based approach, we instead have to work from the cosmological parameters that define the simulations used by the emulator. The comparison between approaches therefore has to be undertaken by comparing posterior surfaces for the cosmological parameters, having matched priors, including the $\theta_{*}$ prior discussed in \cref{sec:theta*}.

\subsection{Baseline $\Lambda \rm{CDM}$} \label{sec:LCDM}

\begin{figure}[!htbp]
    \centering
    \includegraphics[width=\columnwidth]{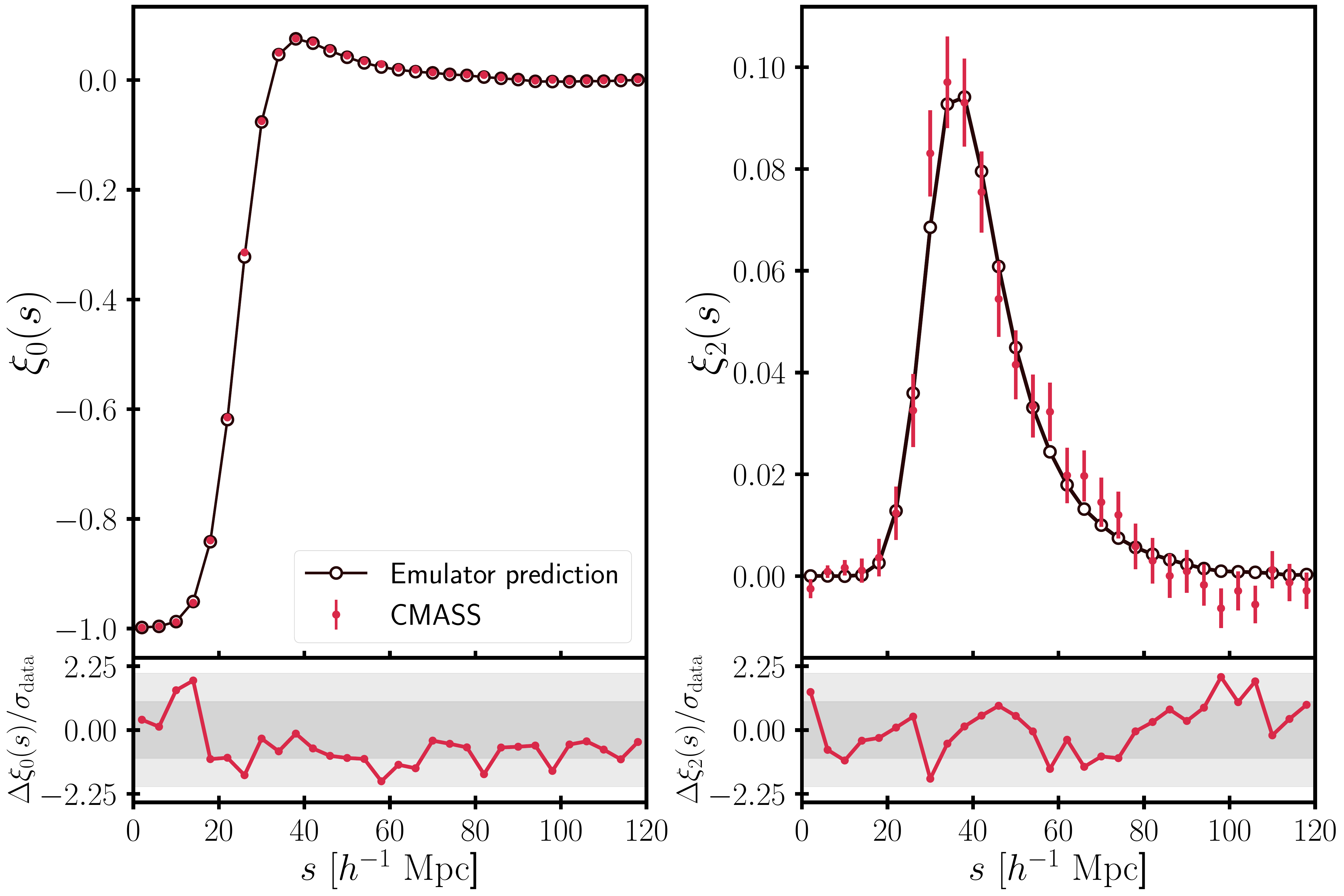}
    \caption{Emulator's fit of the monopole and quadrupole of the void-galaxy cross-correlation function from the CMASS sample (top panels), with the residuals between our fit and the measured monopole and quadrupole (bottom panels). With 1 and 2 $\sigma$ regions shaded in. }
    \label{fig:CMASS_prediction}
\end{figure}

\begin{figure}
    \centering
    \includegraphics[width=0.95\columnwidth]{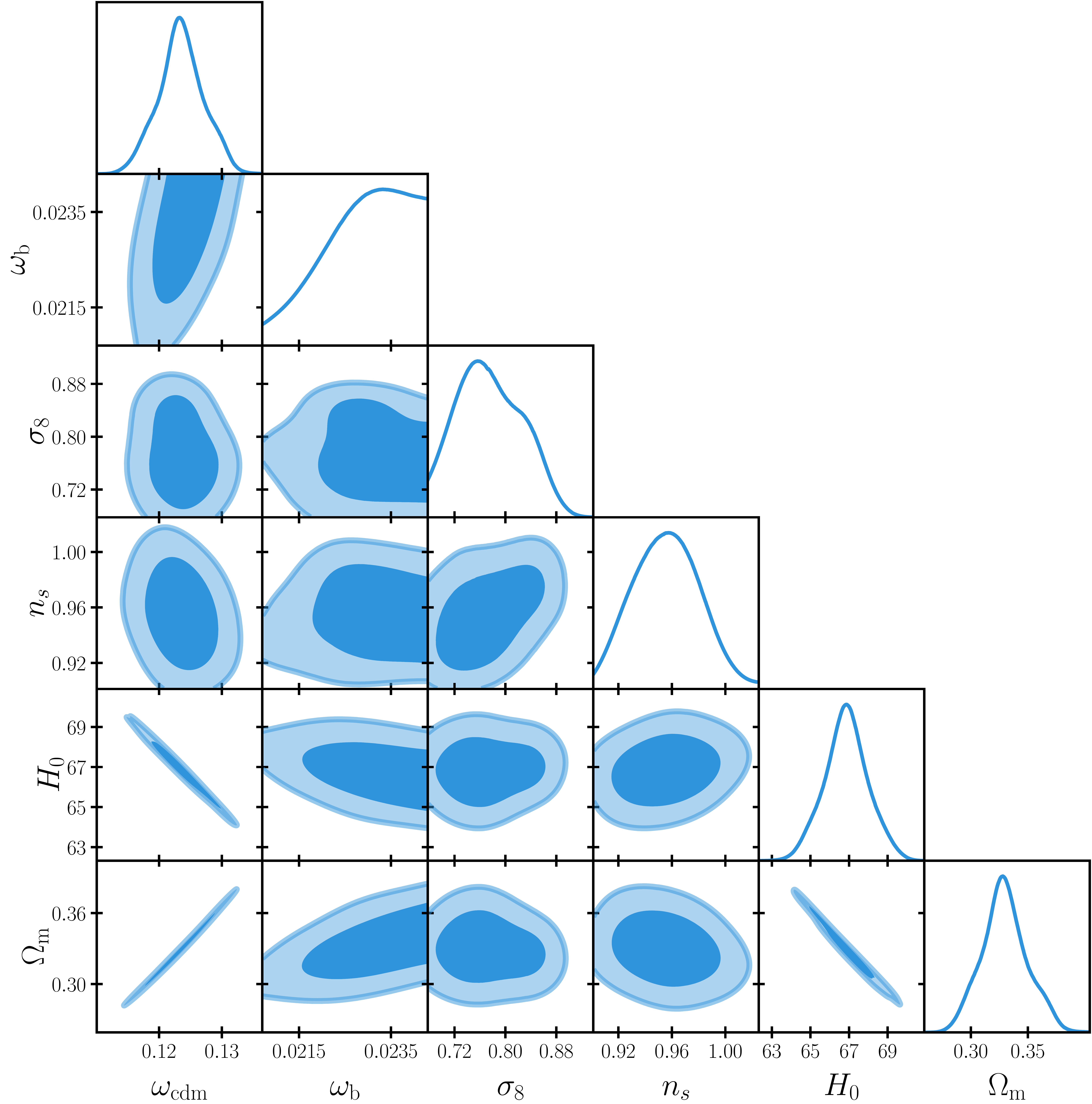}
    \caption{Posterior from fitting CMASS data using the emulator for a flat $\Lambda$CDM model, where $\left\{\omega_{\rm b},\omega_{\rm cdm},\sigma_8,n_{\rm s} \right\}$ are our cosmological free parameters and where we have marginalized over the HOD parameters. $H_0$ is derived from fixing $\theta_{*}$. Consequently, $\Omega_{\rm m}$ can computed from our free parameters $\omega_{\rm cdm}$ and $\omega_{\rm b}$. Fixing $\theta_{*}$ will also influence the baryon fraction, and subsequently $\omega_{\rm b}$. }
    \label{fig:CMASS_cosmoposterior}
\end{figure}

We first consider a flat $\Lambda$CDM model, and fit to the data using the methodology outlined in \cref{sec:mcmc}. The multipole moments of the maximum likelihood solution are compared to the data in \cref{fig:CMASS_prediction}, with the residuals shown in the bottom panel. We find that residuals are not significantly offset at small scales compared to larger scales, which was a result seen in the \abacussummit\ and \nseries\ tests.  We find the greatest deviation in the residuals in the monopole at scales below $20 \Mpch$. While for the quadrupole, we find some deviation at larger scales, but within 2 $\sigma$. The bias is added in quadrature to the covariance to account for it. The projected posterior with cosmological constraints on $\omegac$, $\omegab$, $\sigma_8$, $n_{\rm s}$ (cosmological free parameters) and $\Omega_{\rm m}$, $H_{0}$ is shown in \cref{fig:CMASS_cosmoposterior}. 

\begin{figure}
    \centering
    \includegraphics[width=0.95\columnwidth]{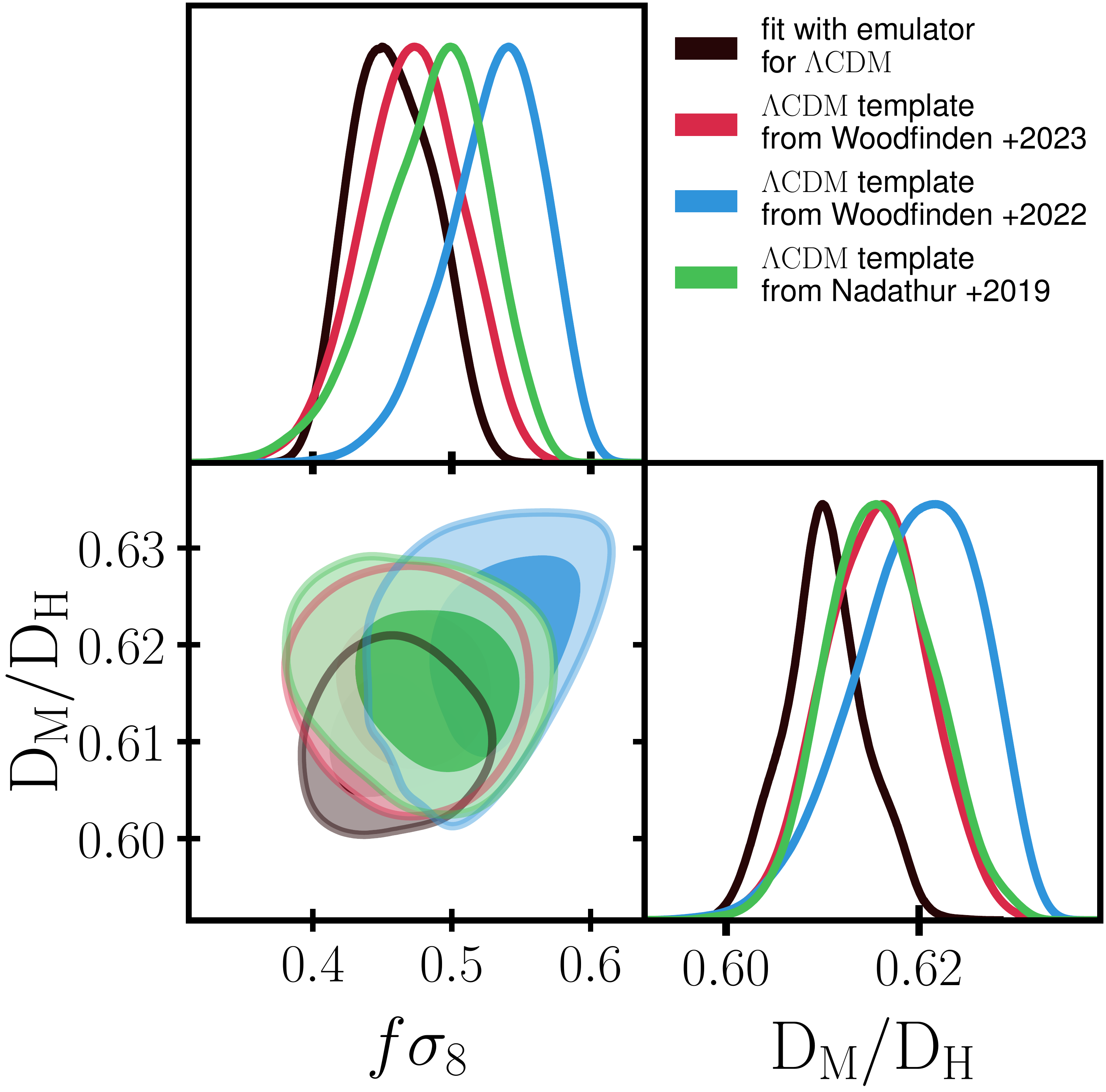}
    \caption{Since we are able to obtain $\Omega_{\rm m}$ due to fixing $\theta_{*}$ we can then derive the Alcock--Paczyński parameter $D_\mathrm{M}/D_\mathrm{H}$ and the structure growth rate $f\sigma_8$ at the effective redshift $z_{\rm eff} = 0.52$. These results (black) can then be compared to the values computed from the template method, using the SDSS sample (red) \cite{Woodfinden_2023}, the CMASS galaxy sample (blue) \cite{Woodfinden_2022} and from an earlier implementation of the template method with CMASS galaxies \cite{Nadathur_2019_CMASS} (green), where the same constraint on $\theta_{*}$ has been included. The templates were fit using the same (narrower) priors as the emulator, which has a minor effect on the constraints produced compared to what is quoted from the respective papers.} 
    \label{fig:CMASS_DMDH}
\end{figure}

\Cref{fig:CMASS_DMDH} shows the 2D posterior for $D_\mathrm{M}/D_\mathrm{H}$ and $f\sigma_8$ derived from, and marginalised over, the $\Lambda \rm{CDM}$ parameters. This is compared to the results reported by \cite{Woodfinden_2022} and \cite{Nadathur_2019_CMASS} using the template method. We compare against the measurements of $D_\mathrm{M}/D_\mathrm{H}$ and $f\sigma_8$ obtained from \cite{Woodfinden_2022} at the $0.4 < z < 0.5$ and $0.5 <z < 0.6$ redshift bins that encompass the CMASS data. We call this sample "CMASS template -- 2 bins". Using \cite{Nadathur_2019_CMASS}, we also compare against previous measurements made assuming a single redshift bin from \cite{Nadathur_2019_CMASS}, which covers a wider redshift range: $0.4 < z < 0.73$.  We call this sample "CMASS template -- 1 bin". We have repeated this procedure including measurements from all five redshift bins of \cite{Woodfinden_2022}, called "SDSS template". Although the bins used in both \cite{Woodfinden_2022} and \cite{Nadathur_2019_CMASS} cover a slightly larger redshift range than used in our analysis, this will not affect our conclusions significantly (see \cref{sec:conclusions}). Since our emulator was constructed using a single time slice, we do not further adjust the redshift range of the data considered in this work, in order to ensure that the emulator predictions remain valid throughout the sample. To match the analysis using the template-based results to those from the emulator, we used \textsc{cobaya} to find parameter constraints for adopting either a $\Lambda$CDM or $w$CDM model given the set of $f\sigma_8$ and $D_\mathrm{M}/D_\mathrm{H}$ measurements and the same priors used in the emulator-based approach, including the $\theta_*$ prior. Narrowing the prior range to match the emulator's does affect the constraints obtained from these template models. We then recomputed derived parameters $D_\mathrm{M}/D_\mathrm{H}$ and $f\sigma_8$ at the effective redshift $z_{\rm{eff}} = 0.52$, within the context of these cosmological models. Note that the derived parameters, even from the template-based approach, will be dependent on the cosmology adopted and, in general, will be more constrained than the direct measurements. Adopting this procedure allows for a fair comparison between template and emulator based approaches, allowing us to compare derived parameter combinations from both methods with similar assumptions. Reference \cite{Hamaus2020:2007.07895} previously considered the choice of fiducial $\Omega_{\rm{m}}$ as a systematic. They found similar constraints suggesting this was either not important, or its effect could be absorbed by their nuisance parameters. They found constraints of $\Omega_{\rm{m}} = 0.312 \pm 0.012$, and $\rm{D_{\rm{M}}}/\rm{D_{\rm{H}}} = 0.588 \pm 0.004$, within $1.5 \sigma$ and $3.1 \sigma$ of ours, respectively. 

\begin{table}[h!]
    \centering 
    \small
    \addtolength{\leftskip}{-2cm}
    \addtolength{\rightskip}{-2cm}
    \renewcommand{\arraystretch}{1.2}
    \begin{tabular}{|c|c|c|c|}\hline
         Result & $\omegac$  & $\sigma_8$  & $w_0$ \\ \hline
         $\Lambda \rm{CDM}$ (emulator) & $0.1235\pm 0.0037$ &  $0.777^{+0.047}_{-0.062}$ & --- \\
         $w \rm{CDM}$ (emulator) & $0.1236^{+0.0062}_{-0.0051}$ &   $0.770^{+0.039}_{-0.056}$ &  $-1.008^{+0.079}_{-0.10}$ \\ \hline
         $\Lambda \rm{CDM}$ (SDSS template \cite{Woodfinden_2022}) & $0.1271\pm 0.0048$ & $0.792^{+0.071}_{-0.058}$ &  --- \\
         $w\rm{CDM}$ (SDSS template \cite{Woodfinden_2022}) & $0.1255^{+0.0089}_{-0.0053}$  & $0.784\pm 0.069$ &  $-0.99\pm 0.15$ \\ \hline
         $\Lambda \rm{CDM}$ (CMASS template -- 2 bins \cite{Woodfinden_2022}) & $0.1305^{+0.0069}_{-0.0039}$ &  $0.877^{+0.066}_{-0.040}$ &  --- \\
         $w \rm{CDM}$ (CMASS template -- 2 bins \cite{Woodfinden_2022}) &$0.1279^{+0.0095}_{-0.0049}$ & $0.874^{+0.075}_{-0.043}$ & $-0.935^{+0.18}_{-0.094}$\\  \hline
         $\Lambda \rm{CDM}$ (CMASS template -- 1 bin \cite{Nadathur_2019_CMASS})  & $ 0.1273\pm 0.0047$ & $0.814^{+0.082}_{-0.055}$ & --- \\
         $ w\rm{CDM}$ (CMASS template -- 1 bin \cite{Nadathur_2019_CMASS}) &  $0.1243^{+0.0087}_{-0.0060}$ & $0.802^{+0.089}_{-0.064}$ & $-0.96^{+0.19}_{-0.11}$\\ \hline
    \end{tabular}
    \caption{Table of credible intervals for cosmological parameters from various fits. "CMASS template -- 2 bins" describes the constraints obtained with a template using the two redshift bins of \cite{Woodfinden_2022} covering most of the CMASS galaxies: $0.4 < z < 0.5$ and $0.5 < z < 0.6$. "CMASS template -- 1 bin" indicates the results from applying the template method using the CMASS sample to a single redshift bin $0.4 < z < 0.73$ from \cite{Nadathur_2019_CMASS}, while "SDSS template" results include all 5 redshift bins. Note that 2D posteriors for both fits hit the prior walls, defined by the emulator, for $\omegab$ and $\sigma_8$.}
    \label{tab:cosmo_result_table}
\end{table}

We report the results from the template method in comparison to our emulator in \cref{tab:cosmo_result_table,tab:derived_result_table}. 
We see a reduction in errors with the emulator method compared to the template-based approach for both the cosmological parameters, $\Omega_{\rm{m}}$, $\sigma_8$ and the derived AP and RSD parameters $D_\mathrm{M}/D_\mathrm{H}$ and $f\sigma_8$. Specifically, we see at least a $25\%$ reduction in errors for the AP parameter, and a $8.4\%$ reduction in errors for the RSD parameter depending on the CMASS sample used. These arise from at least a $28\%$ reduction in errors for $\Omega_{\rm{m}}$ and at least a $1.7\%$ reduction in errors for $\sigma_8$. Because we have added the $\theta_{*}$ prior to the template results, this is a clear indication that the improvement in constraining power does not only come from this prior, but also from the full-shape of the VG-CCF, alongside a proper treatment of the Alcock--Paczyński effect in terms of modelling the void-galaxy cross-correlation function, and the adoption of an emulator model over a simulation-based template model for velocity fields around voids. This methodology produces 1D marginalized constraints on $\Omega_{\rm{m}} =0.330\pm 0.020$ and $\sigma_8 =  0.777^{+0.047}_{-0.062}$, which we highlight as the main result of this work. Additionally, the emulator produces a constraint on $n_{\rm s} = 0.956\pm 0.024$.

\begin{table}[!h]
    \centering 
    \small
    \addtolength{\leftskip}{-2cm}
    \addtolength{\rightskip}{-2cm}
    \tabcolsep=2.5pt
    \renewcommand{\arraystretch}{1.2}
    \begin{tabular}{|c|c|c|c|c|}
    \hline
         Result & $D_\mathrm{M}/D_\mathrm{H}$ & $f\sigma_{8}$ & $\Omega_{\rm m}$ &  $\sigma_8$ \\ \hline

         $\Lambda$CDM (emulator) & $0.6103\pm 0.0042$  & $0.458^{+0.029}_{-0.033}$ & $0.330\pm 0.020$ & $0.777^{+0.047}_{-0.062}$ \\ \hline
         $w\rm{CDM}$ (emulator) &  $0.6089^{+0.0061}_{-0.0073}$ & $0.463\pm 0.033$ & $0.329^{+0.019}_{-0.024}$ & $0.778^{+0.048}_{-0.059}$ \\ 
         \thickhline
         $\Lambda$CDM (SDSS template \cite{Woodfinden_2022}) & $0.6157\pm 0.0055$ & $0.472^{+0.041}_{-0.034}$ & $0.357^{+0.026}_{-0.029}$ & $0.792^{+0.071}_{-0.058}$\\ \hline
         $w\rm{CDM}$ (SDSS template \cite{Woodfinden_2022}) & $0.6159\pm 0.0056$& $0.473\pm 0.037$ & $0.348^{+0.049}_{-0.034}$ & $0.784\pm 0.069$\\ \hline
         $\Lambda$CDM (CMASS template -- 2 bins \cite{Woodfinden_2022}) & $0.6199^{+0.0080}_{-0.0051}$ & $0.529^{+0.039}_{-0.029}$ & $0.378^{+0.040}_{-0.028}$ & $0.877^{+0.066}_{-0.040}$\\ \hline

         $w$CDM (CMASS template -- 2 bins \cite{Woodfinden_2022}) & $ 0.6210\pm 0.0069$ & $0.523^{+0.049}_{-0.033}$ & $0.363^{+0.055}_{-0.033}$ & $ 0.874^{+0.075}_{-0.043}$\\ \hline

         $\Lambda \rm{CDM}$ (CMASS template -- 1 bin \cite{Nadathur_2019_CMASS}) & $0.6160\pm 0.0055$ & $0.486^{+0.046}_{-0.033}$ & $0.358^{+0.025}_{-0.030}$ & $0.814^{+0.082}_{-0.055}$ \\ \hline
         $w\rm{CDM}$ (CMASS template -- 1 bin \cite{Nadathur_2019_CMASS}) & $0.6159\pm 0.0055$ & $0.487^{+0.050}_{-0.034}$ & $0.341^{+0.047}_{-0.038}$ & $0.802^{+0.089}_{-0.064}$ \\ \hline
         Density Split (emulator \cite{paillas2023cosmological}) & ---  &  $0.462 \pm 0.020$ & $0.306 \pm 0.012$ &  $0.792 \pm 0.034$\\ \hline
         Galaxy 2PCF (emulator \cite{Yuan_2022}) &  --- & $0.444 \pm 0.016$ & $0.309^{+0.018}_{-0.017}$ & $0.762 \pm 0.024$\\ \hline
         WST BOSS (emulator \cite{valogiannis2023precise}) & --- & $0.469 \pm 0.012$ & $0.322 \pm 0.007$ & $0.834^{+ 0.058}_{-0.039}$ \\ \hline
    \end{tabular}
    \caption{Table of derived parameters and $\sigma_8$, with their means and standard deviations shown. The density-split clustering and the galaxy two-point correlation function (2PCF) emulator were fitted to the same data, the BOSS CMASS NGC catalog with similar redshift binning: $0.45 \leq z \leq 0.60$ for Density Split  and  $0.46 < z < 0.60 $ for the 2PCF emulator. The wavelet scattering transform (WST) analysis was run across a redshift range of $0.46 < z < 0.57$, with $z_{\rm{eff}}=0.515$ used for the $f(z)\sigma_8(z)$ calculation. }
    \label{tab:derived_result_table}
\end{table}

The cosmological parameters are given in \cref{tab:cosmo_result_table}, while the derived parameters are presented in \cref{tab:derived_result_table}. We also compare the credible intervals obtained from the posteriors to those derived using the template method. A similar improvement in constraining power is seen for both the AP and RSD parameters, and for $\Omega_{\rm m}$ and $\sigma_8$. This further illustrates that a full-shape analysis of the VG-CCF can produce unbiased and competitive constraints on $\Omega_{\rm{m}}$ and $\sigma_8$.

\subsection{Extending to $w\rm{CDM}$ cosmologies}

\begin{figure}
    \centering
    \includegraphics[width=0.95\columnwidth]{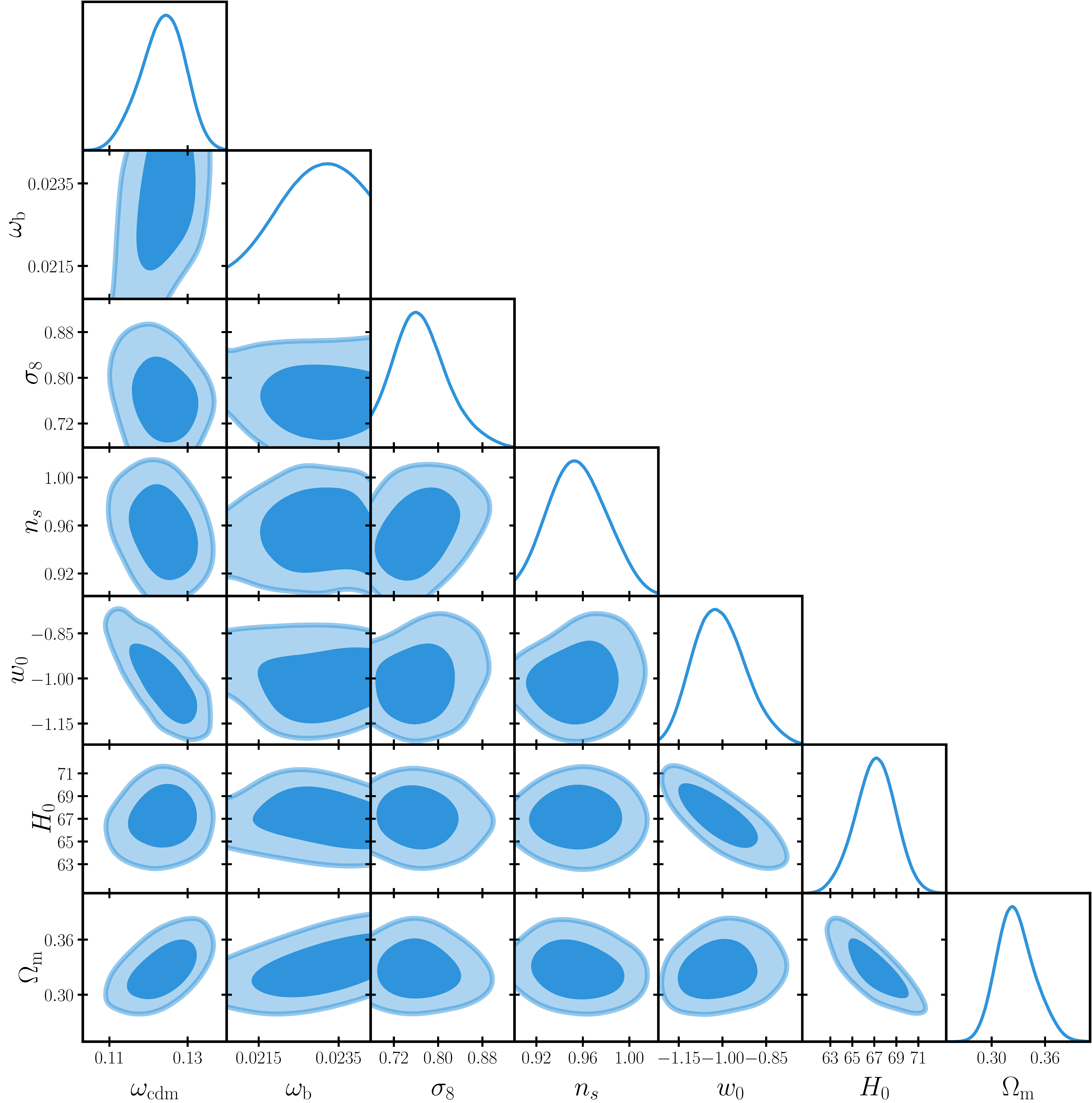}
    \caption{Posterior from fitting CMASS data using the emulator for a flat $w \rm{CDM}$ model, where $\left\{\omega_{\rm b}, \omega_{\rm cdm}, \sigma_8, n_{\rm s}, w_0 \right\}$ are our cosmological free parameters and where we have marginalized over the HOD parameters. $H_0$ is derived from fixing $\theta_{*}$. Consequently, $\Omega_{\rm m}$ can computed from our free parameters $\omega_{\rm cdm}$ and $\omega_{\rm b}$. Fixing $\theta_{*}$ will also influence the baryon fraction, and subsequently $\omega_{\rm b}$. }
    \label{fig:CMASS_cosmoposterior_WCDM}
\end{figure}

\begin{figure}
    \centering
    \includegraphics[width=0.95\columnwidth]{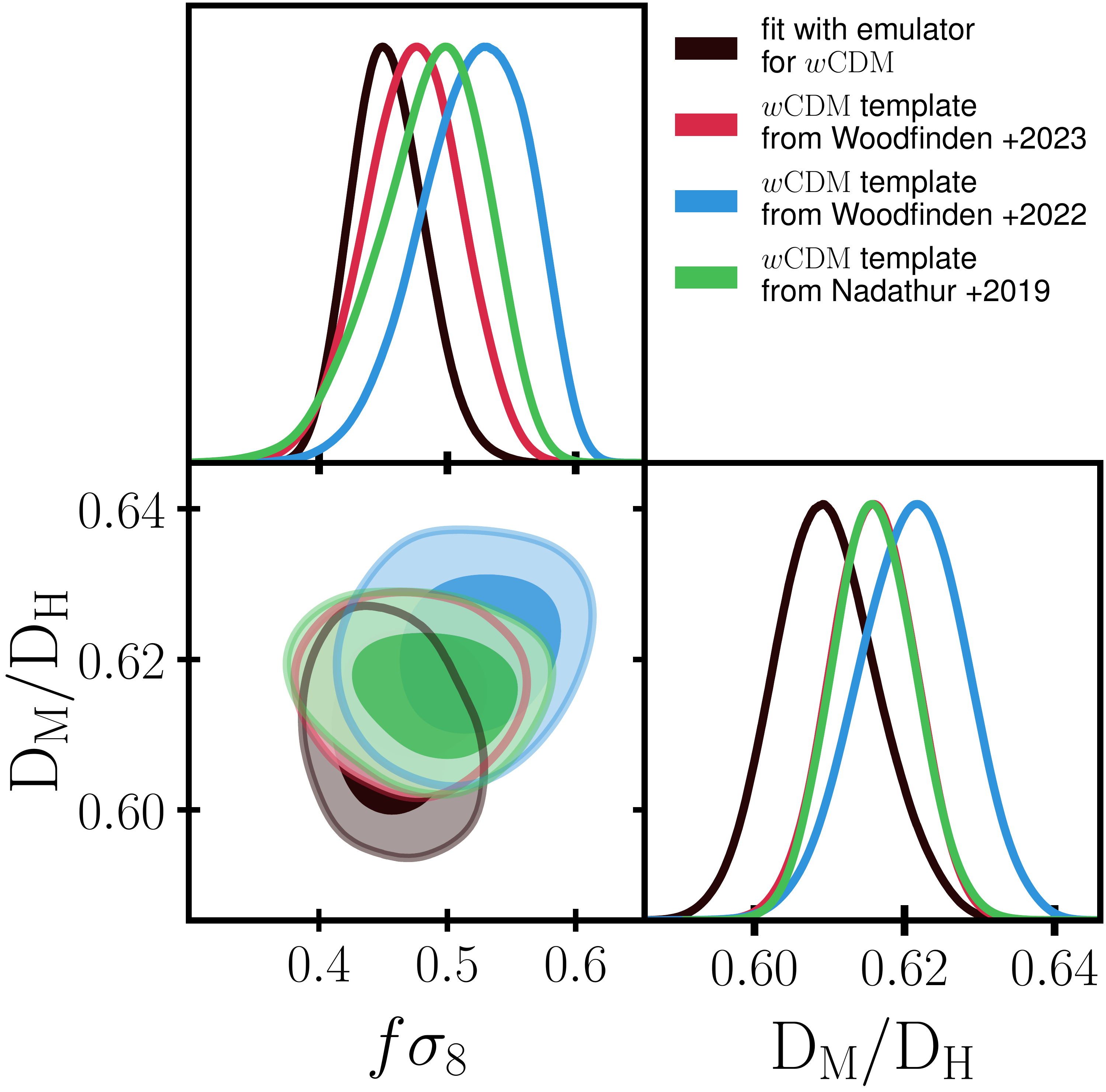}
    \caption{We again derive the Alcock--Paczyński parameter $D_\mathrm{M}/D_\mathrm{H}$ and the structure growth rate $f\sigma_8$ from our emulator's free parameters and $\Omega_{\rm{m}}$, this time for $w\rm{CDM}$ cosmologies. These values (black) can then be compared to the results of using the template method, as applied to the SDSS sample (red) by \cite{Woodfinden_2023},  using a subset of the CMASS galaxy sample (blue) by \cite{Woodfinden_2022} and an earlier analysis of CMASS galaxies (green) by \cite{Nadathur_2019_CMASS} where in all of these templates the same constraint on $\theta_{*}$ has been included.} 
    \label{fig:CMASS_DMDH_Wcdm}
\end{figure}

We extend our analysis by considering a $w\rm{CDM}$ cosmology where the dark energy equation of state $w$ is constant in time, but allowed to have $w\ne1$. The 2D posterior of the $w\rm{CDM}$ parameters can be found in \cref{fig:CMASS_cosmoposterior_WCDM} , where we have marginalized over the HOD parameters. We note significant improvements over the template method in the constraining power of $\Omega_{\rm{m}}$, $\sigma_8$ and $w$ with the emulator model, as seen in \cref{fig:CMASS_DMDH_Wcdm}. Our emulator model favours lower values of $\sigma_8$ and $\Omega_{\rm{m}}$ when compared to the template model. The results are presented in \cref{tab:cosmo_result_table} \& \cref{tab:derived_result_table}. When considering $w\rm{CDM}$ models, we end up with a significantly wider posterior for $D_\mathrm{M}/D_\mathrm{H}$ than for $\Lambda$CDM models, with the upper $1\sigma$ bound enlarged by a factor of 1.5, and the lower 1$\sigma$ bound enlarged by a factor of 1.7. We also find a $0.3\sigma$ shift in the mean of the posterior towards a smaller $D_\mathrm{M}/D_\mathrm{H}$ and a $0.2\sigma$ increase in the mean $f\sigma_8$, with the posterior widths remaining similar. The means and standard deviations in $\Omega_{\rm{m}}$ and $\sigma_8$ are largely unchanged.

\subsection{Testing the parameters of the fits}

\begin{figure}
    \centering
    \includegraphics[height=15.5cm]{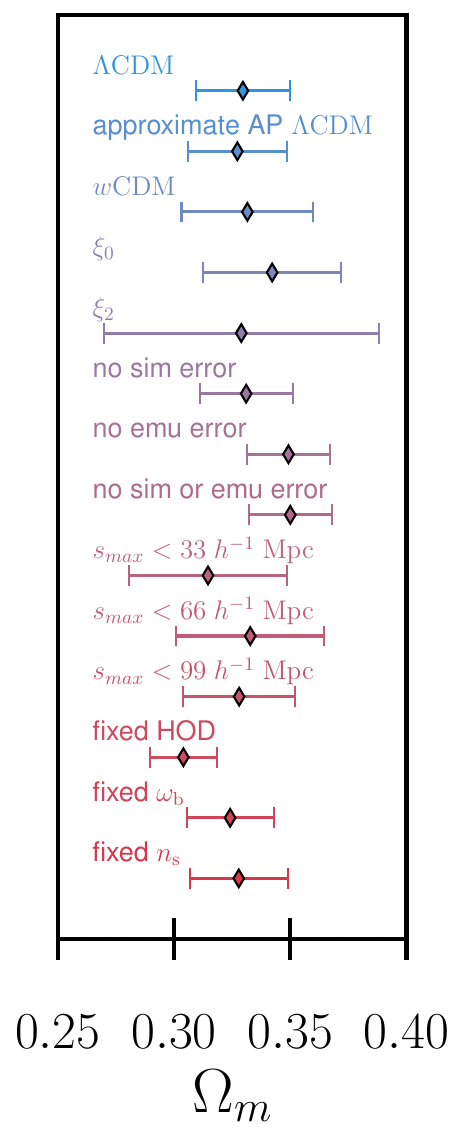}
    \hspace{1cm}
    \includegraphics[height=15.5cm]{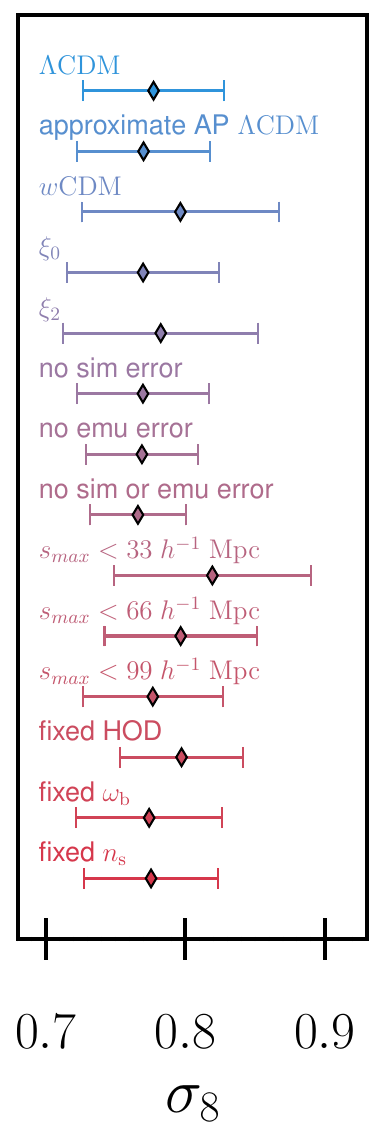}
    \caption{Marginalized constraints on $\Omega_{\rm m}$ (left) and $\sigma_8$ (right) from the BOSS CMASS data, under different settings for the emulator. We have explored extensions to flat $\Lambda$CDM, different implementations of the Alcock--Paczyński effect to the void-finder, fitting with either the monopole or quadrupole only, as well as the effects of removing different parts of our covariance or limiting the scale of emulation $s_{\rm max}$, as well as fixing $\omega_{\rm b},n_{\rm s}$ and the HOD parameters. }
    \label{fig:whisker_om_s8}
\end{figure}

\Cref{fig:whisker_om_s8} shows the change in constraints on the primary cosmological parameters, $\Omega_{\rm{m}}$ and $\sigma_8$, constrained by the voids, after we make a number of changes to the method. After providing the baseline $\Lambda$CDM and $w\rm{CDM}$ results, the rows show the results of using a simplified model of the AP effect in the void-galaxy cross-correlation function matching that commonly seen in the literature where void centres are assumed to be shifted by AP as with galaxies, fitting to the  monopole or quadrupole only, and reducing the covariance by removing components as described in \cref{sec:covariance}. We also consider reducing the scales fitted, not varying the HOD parameters, and fitting reduced sets of cosmological models, and fixing either $\omegab$ or $n_s$ to their Planck 2018 values \citep{Planck2018Cosmo}. We find our results are remarkably robust to all of the changes considered.

The monopole provides more information than the quadrupole for $\Omega_{\rm m}$, indicating that the anisotropic stretching of the cross-correlations is not as important as the spherically averaged shape for this parameter. The result is more balanced for $\sigma_8$. Our interpretation of this is that the $\sigma_8$ constraint is limited by the RSD measurement, while the dependence of $\Omega_{\rm m}$ on the AP effect is more nuanced. This matches the observation that, for $\Omega_{\rm m}$ unlike $\sigma_8$, there remains significant information on large scales, suggesting information is being extracted from the cosmological large-scale clustering. Also, we would expect that fixing the HOD or alternative cosmological parameters increases the amount of signal from the shape of clustering, and this also improves the $\Omega_{\rm{m}}$ error, but has a weaker effect on $\sigma_8$. 

The sub-covariances discussed in \cref{sec:covariance} arising from the data and emulator add roughly equally to the combined covariance, suggesting a matching between the current size of the \abacussummit\ simulation suite and the data. For future surveys where the data covariance is improved, emulators with a denser sampling of parameters would be required to reduce the errors associated with the emulators, such that they are sub-dominant. 

\section{Discussion} \label{sec:discussion}

The results presented in \cref{sec:Results-data} rely on several underlying decisions for our methodology, particularly the choice of voids to include, priors, and scales fitted, which all have a downstream effect on the constraining power arising from the monopole and quadrupole. To compare with the work of \cite{Woodfinden_2022}, we used a similar treatment for our void catalogs, stacking all voids with radius $>20$h$^{-1}$Mpc for subsequent analysis. Note that the voids are not the same, as the emulator method uses voids found in redshift space, while \cite{Woodfinden_2022} uses voids found post-reconstruction (as described in \cref{sec:intro}). We also use the {\tt voxel} void-finding routine, while \cite{Woodfinden_2022} uses {\tt ZOBOV} \citep{Neyrinck2008:0712.3049}, although figure~11 of \cite{Radinovic_inprep} shows that the voids found are similar for both approaches.  An important caveat is that the emulator's parameter space is fundamentally limited by the mocks used in training. Predictions made by the emulator outside of the range of the \abacussummit\ mocks cannot be used in our model. This informs our choice of priors when comparing with the template model, where we have to narrow our prior ranges used in \textsc{cobaya} to match the \abacussummit\ prior range.

In addition, our training data should mimic all aspects of the observations. In the analysis presented, we instead use simplified mocks, testing the robustness to observational effects. We believe that the tests presented justify these simplifications. However, further testing and adopting training mocks using full lightcones with the survey geometry would make the method more robust for upcoming analyses. In that sense, our analysis should be considered a first step along this path.

One of the effects that would be captured in lightcones and our cut-sky Nseries tests is the redshift evolution of the AP effect. Our training data had a single AP stretch parameter, computed for the box redshift at $z=0.5$. Whereas the AP shifts varies with redshift. In the case of a serious mismatch between the fiducial and the true (box) cosmology, this could cause a mismatch in the average behaviour of void-galaxy pairs. To test for this, we averaged the AP shifts for a random set of void-galaxy pairs from $0.45<z<0.60$, and compared this with the expected shift at $z=0.5$, corresponding to the box redshift. For 10 different cosmologies chosen within the $2\sigma$ of our $\omegac$ and $\sigma_{8}$ measurements. We find that the average vs. expected AP shifts differ by $0.15\%$ in $q_{\parallel}$ and $0.11\%$ in $q_{\perp}$ at most. These are significantly below our errors---our measurement of the AP factor has an error of $0.7\%$ for $\Lambda\rm{CDM}$ models. Larger survey volumes and training sets will necessitate using lightcone mocks as the errors in $D_\mathrm{M}/D_\mathrm{H}$ will shrink.

\subsection{Impact of the $\theta_{*}$ prior}

As discussed, the \abacussummit\ simulations were used to train our emulator and have the same $\theta_{*}$. This limits our analysis to only consider models with the same projected CMB peak position. In order to be self consistent, and not use the emulator to predict models with very different $\theta_{*}$ we must either only consider constraints including additional CMB data, or include a prior as we have done here. We chose the latter route in order to limit the information to voids as much as possible. While theoretically justified as the CMB observations are very strong \cite{Planck2018Cosmo}, it would have been simpler in terms of interpretation were this constraint not present in the training sample of mock catalogues.

For a $\Lambda$CDM model at fixed physical baryon density the $\theta_{*}$ prior is equivalent to a prior on $\Omega_{\rm m}h^{3.4}$ \cite{Percival_2002}. The dependence on the baryon fraction is weak, such that even broad bounds on $\omega_{\rm b}$, as given in \cref{tab:priors}, lead to only a small weakening of this degeneracy. While the primary information coming from the voids remains the AP and RSD parameters, the emulator also recovers information from the shape of the clustering signal. This information, together with the $\theta_{*}$ prior explains why our $\Lambda$CDM fit is able to constrain $H_0$ and $\sigma_8$ (or $\log A$), as seen in \cref{fig:CMASS_DMDH}, for example, breaking a number of degeneracies present in previous void only analyses.  This is underscored by the marginalized constraints shown in \cref{fig:whisker_om_s8}, where the monopole and quadrupole both contribute to constraining $\Omega_{\rm{m}}$ and $\sigma_8$.

To compare to the results from \cite{Woodfinden_2022, Woodfinden_2023}, we must add the same $\theta_{*}$ prior to their results, and can only compare within the set of cosmological parameters for a given model, rather than the measurements directly, as described in \cref{sec:LCDM}. The interpretation of the cosmological constraints is complicated by having to disentangle the measurements from this prior.

\begin{figure}
    \centering
    \includegraphics[width=0.9\columnwidth]{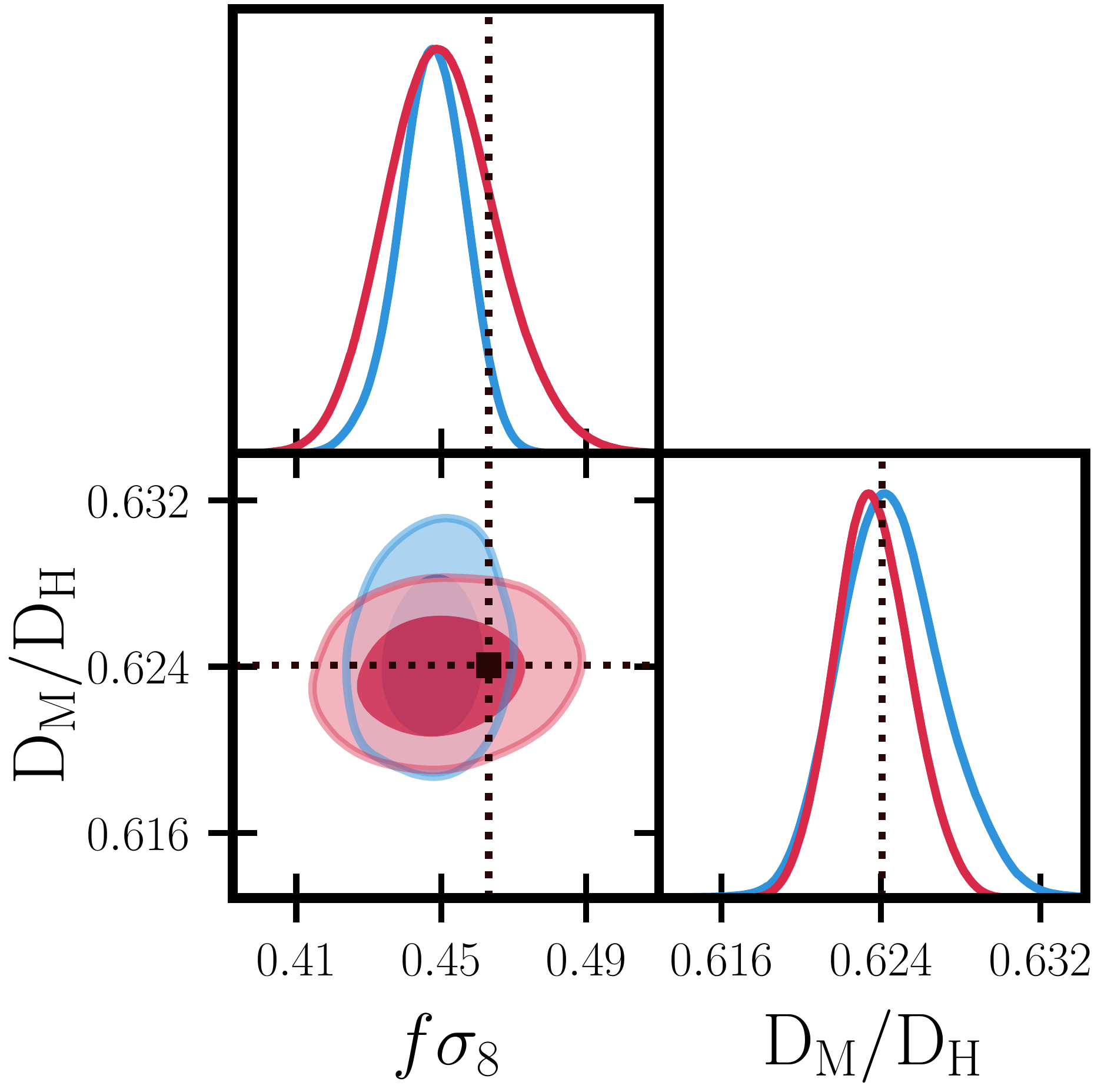}
    \caption{A comparison of the constraints obtained on $f\sigma_8$ and $D_\mathrm{M}/D_\mathrm{H}$ through two sets of training data obtained generated with two different assumptions for the  Alcock--Paczyński (AP) effect applied to voids, one where the AP effect is applied after void finding (red), and the better match to the process using when analysing real data, where the AP effect is applied before void finding (blue). This figure was generated by fitting a mock with the same HOD within the {\tt c002} \abacussummit\ non-fiducial cosmology by emulators trained with the different methods.
    }
    \label{fig:AP_true_approx_Abacus}
\end{figure}

\subsection{The Alcock--Paczyński effect}

The emulator method easily allows the AP effect on void finding and on the void-galaxy cross-correlations to be fully included by shifting the galaxies before void finding, as would be the case for an analysis of data. To test this, we emulate the monopole and quadrupole moments of the redshift-space void-galaxy cross-correlation, with the void-finder obtaining void-centres under the following two scenarios: i) Voids are found in the true cosmology of the training data and the void centres and galaxies are shifted by the AP stretching. ii) The AP shift is applied to galaxies before the void centres are found, a scenario matching the process applied to data. (i) is an assumption commonly used in the literature (e.g. \cite{Woodfinden_2022}) as it simplifies the analysis. 

If the fiducial cosmology exactly matched the truth, we would not expect a systematic offset in the best-fit position for (i) versus (ii), but would still see a change in the derived credible intervals. \Cref{fig:AP_true_approx_Abacus} shows how our results actually change between these two scenarios. Given that the fiducial cosmology adopted lies near the centre of the credible intervals derived from our fits, as shown in \cref{fig:AP_true_approx_Abacus}, it is therefore expected that the maximum posteriors of the results from different approaches are close. For (i), we see a improvement in the constraining power of $D_\mathrm{M}/D_\mathrm{H}$, while $f\sigma_8$ is less well constrained. This suggests that the simplified approach previously used was underestimating the errors in $D_\mathrm{M}/D_\mathrm{H}$.

The change in the sizes of the credible intervals is significant, and is discussed further in \cite{Radinovic_inprep}. When we model the AP effect in the void-galaxy cross-correlation function from the same simplistic base assumptions of previous studies it does change the emulator-based models as the AP deviation becomes stronger, reducing the overall impact of the AP effect and leading to wider credible intervals. \cite{Radinovic_inprep} offers a thorough exploration of the assumption and how its effect translates into differences in the void-galaxy cross-correlation function. They show that this is a fundamental consequence of the way that the AP effect affects void finding, and not a consequence of the emulator-based approach. Our analysis agrees with \cite{Radinovic_inprep}'s conclusion that void-centres' response to the AP distortions differs significantly from that of galaxies. While our results agree with the explanation provided by \cite{Radinovic_inprep}, they do not by themselves prove this assertion. The emulator method does improve a common issue in void analyses: the cosmology being tested will change the topology of the density field and thus the distribution of spurious voids. As voids are found in these shifted catalogues in the emulator method, exactly as they would be in the data as if the cosmology was correct, then the number of spurious voids expected in the data matches that in the model. The only caveat to this is that our emulator is built on simulation cubes and not cutsky mocks, but we have tested the impact of this.

\subsection{Comparison with other simulation-based clustering analyses}

Previous works have also used mock galaxy catalogues to train simulation-based models of various clustering statistics. In particular, among those works who have used \abacussummit\ to calibrate their models and applied to BOSS data, \cite{Yuan_2022} emulated the small-scale galaxy two-point correlation function to obtain constraints on the halo occupation and $\Lambda$CDM parameters. \cite{paillas2023cosmological} emulated density-split clustering statistics to obtain constraints on $\Lambda$CDM and single-parameter extensions, including $N_{\rm eff}$ and $w_0$. Similarly, \cite{valogiannis2023precise} emulated wavelet scattering transforms (WST) to obtain cosmological constraints on the base $\Lambda$CDM parameters and its extensions. These constraints are summarized in \cref{tab:derived_result_table}. Additionally, beyond two-point clustering with voids has been used to enhance cosmological constraints, notably \cite{thiele2023} used the void-size function,  the void galaxy cross power spectrum, and the galaxy auto power spectrum to constrain neutrinos masses in BOSS. Our work and  \cite{valogiannis2023precise,paillas2023cosmological,thiele2023} show that going beyond the two-point clustering can lead to significant improvements in constraining power. An important fraction of this constraining power is found at smaller scales, $s <20 \Mpch$. The error of our emulator is more dominant at small scales, where this signal is coming from, highlighting the importance of further improvements in emulation techniques and sets of training data.

Characterizing the galaxy-halo connection in greater detail would also be another crucial ingredient in this methodology --- we used a simple HOD model, without characterizing assembly bias and marginalizing over baryonic effects and velocity bias. The treatment of these effects might prove important in future work; the work of \cite{storeyfisher2022aemulus} with the marked power spectrum and probability underdensity function found that degeneracies between halo bias parameters and cosmological parameters can be broken by alternative clustering statistics. This potential is noted by \cite{paillas2023cosmological} with a more detailed HOD model using the CMASS data. This motivates exploring if the shape of the VG-CCF can produce similar degeneracy breaking, either under a more detailed HOD model, or with different galaxy-halo connection models incorporated into the training (e.g. SHAM mocks). Another question raised is how the shot noise from our voxel voids might limit these constraints, in comparison to other low-density summary statistics.

There is significant information to be gained beyond 2-point statistics, in the form of the initial information suppressed by non-linear evolution, and from projection effects \cite{Samushia_2021}. Each method discussed above can be considered a compression of this information, and will be sensitive to different cosmological processes, and hence parameter combinations \cite{Beyond2ptcollaboration2024parametermasked}. Using voids, we obviously limit the range of densities considered and focus on AP and RSD information. While we therefore provide less information overall, using only voids has the benefit of limiting the reliance on correctly modelling the full non-linear behaviour of the galaxy field at high overdensities.

\subsection{Future improvements in the method}

Given the forthcoming improvements in available spectroscopic galaxy samples from DESI \cite{DESI2016a,DESI2016b} and Euclid \cite{Laureijs_2011} we believe that the work presented here will be of increasing importance in the future. In particular, the emulator-based approach allows us to use self-consistent profiles for the voids we are fitting, and allows a full inclusion of the AP effect. We have shown that both only have a small effect on the data analysed here, as demonstrated by the match between our results and those of \cite{Woodfinden_2022}. However, these differences will become more important for future data. 

Optimizations in the void finder could have significant downstream effects on improving the accuracy of the emulator, as the largest bottleneck in an emulator-based model stems from the $N$-body simulations used for training data. Improvements in the speed and efficiency of the void-finder could lead to significant improvements in coverage of HOD parameter space allowing us to generate a larger training data set, potentially allowing a more thorough exploration of more complex galaxy-halo connection models and the potential constraining power offered by summary statistics exploring low density regions --- as identified in \cite{storeyfisher2022aemulus,paillas2023cosmological}, while potentially reducing the amplitude of systematic bias coming from our emulator model. So far, we have found that while there are detectable biases in the emulation of the VG-CCF at scales below $20-40 \Mpch$, their effects on cosmological parameter recovery are negligible. The fact that the posteriors do not reveal significant differences suggest that the form of the offsets observed does not match that expected for any cosmological model we have tested.

As well as the improvements in the accuracy and number of training simulations discussed in \cref{subsection:abacusmocks}, for the void analysis, one could consider using different void-radius bins, to further extend our data vector and explore the role of void-radius in the VG-CCF. By doing this, we could increase the extra information coming from the shape of the VG-CCF, which we can include with our emulator method. Optimizations in the radius cuts---identifying which scales contribute to constraining $\Omega_{\rm m}$ and $\sigma_8$ would be valuable, allowing an optimal weighting for voids to be included.

We note that our summary statistic is determined by stacking voids of different sizes. Consequently, differences between voids of different sizes would be smoothed out, potentially obscuring the information obtained from the void-galaxy cross-correlation function. This may, however, remove potential dependence on survey specific effects and limitations in the emulator. We leave an exploration of this, which may lead to better constraints to future work. This might include weighting voids of different sizes and/or using thinner bins in radii.

Another limitation of our current emulator is that it is trained on simulations at a single snapshot. This restricts our analysis to a relatively narrow redshift range, where the evolution of the void clustering statistics can be approximated as a constant. In future work, we plan on exploring the addition of redshift evolution to our model using the \abacussummit\ lightcone simulations \cite{Hadzhiyska2022:2110.11413}, which became available during the course of this work.

\section{Conclusions} \label{sec:conclusions}

We have presented a new method for obtaining cosmological information from the clustering of galaxies around voids:
\begin{itemize}
    \item We have trained a neural-network emulator to emulate the void-galaxy cross-correlation function using the redshift-space positions of galaxies. This is the first emulation of the full-shape of the void-galaxy cross-correlation function. We have outlined the methodology used to build this emulator, and used it to produce robust, unbiased cosmological constraints from the CMASS galaxy sample $0.45 < z < 0.6$.

    \item We implemented the Alcock--Paczyński effect in our method, by training our emulator using stretched galaxy catalogs --- emulating the cross-correlations as if the data were analysed using a fiducial cosmology. Consequently, the AP effect was applied to our model, without the simplifications seen in previous studies (e.g. \cite{Woodfinden_2022}). Following \cite{Radinovic_inprep}, we would expect that applying the simplified model of the AP effect on the void-galaxy cross-correlation function would lead to an underestimation of the errors of $\rm{D_M}/ \rm{D_H}$. Validating this conclusion for our analysis would require the use of lightcones to fully avoid the stepwise growth rate expected in cut-sky mocks. We do, however, expect this effect to be small.

    \item The emulator improves the constraining power relative to methods using templates to model the void-galaxy cross-correlation function. This suggest that the full-shape of the void-galaxy cross-correlation function contains significant information that can be used to further enhance the constraining power from voids.

    \item Finally, we note that this work adds to a growing chorus in the literature emphasising the importance of providing mock samples that can be used to train emulators---as upcoming surveys shrink the covariance from our data, it is crucial that emulator bias contributions to the systematic errors become sub-dominant.
\end{itemize}

\section{Acknowledgements}

The authors thank Simone Paradiso for helpful discussions throughout the course of this work, and for his assistance with the \textsc{cobaya} likelihood code. The authors also want to thank Alex Woodfinden for helpful discussions about the template method and its nuances.

This research was enabled in part by the support provided by Compute Ontario (computeontario.ca) and the Digital Research Alliance of Canada (alliancecan.ca). WP acknowledges the support of the Natural Sciences and Engineering Research Council of Canada (NSERC), [funding reference number RGPIN-2019-03908] and from the Canadian Space Agency. SN acknowledges support from an STFC Ernest Rutherford Fellowship, grant reference ST/T005009/2.

The massive production of all MultiDark-Patchy mocks for the BOSS Final Data Release has been performed at the BSC Marenostrum supercomputer, the Hydra cluster at the Instituto de Fısica Teorica UAM/CSIC, and NERSC at the Lawrence Berkeley National Laboratory. We acknowledge support from the Spanish MICINNs Consolider-Ingenio 2010 Programme under grant MultiDark CSD2009-00064, MINECO Centro de Excelencia Severo Ochoa Programme under grant SEV- 2012-0249, and grant AYA2014-60641-C2-1-P. The MultiDark-Patchy mocks was an effort led from the IFT UAM-CSIC by F. Prada’s group (C.-H. Chuang, S. Rodriguez-Torres and C. Scoccola) in collaboration with C. Zhao (Tsinghua U.), F.-S. Kitaura (AIP), A. Klypin (NMSU), G. Yepes (UAM), and the BOSS galaxy clustering working group.
\appendix
\section{Full posteriors for test cosmologies}\label{sec:Appendix}
We have enclosed the full posterior distributions for our six test cosmologies used for evaluating the quality of the emulator's predictions. We find that overall the posterior range for the HOD parameters cosnsitently hits the prior walls, and are uncorrelated with the cosmological parameters. We marginalize over the HOD parameters for our cosmological analysis, from \cref{fig:AbacusTestPosterior}.
\begin{figure}[h!]
    \centering
    \includegraphics[width=\columnwidth]{c000_fullpost_new.pdf}
    \caption{Full posterior of the {\tt c000} test cosmology. True values for the parameters are shown by the solid black lines. Overall, we find that the posterior range for the HOD parameters $\log M_{\rm cut}$, $\log M_1$, $\log\sigma$, $\alpha$, $\kappa$ consistently hit the limits of our priors and are uncorrelated with our cosmological parameters. We then marginalize over these HOD parameters for the remainder of our analysis. }
    \label{fig:c000_full_postApp}
\end{figure}
\begin{figure}[h!]
    \centering
    \includegraphics[width=\columnwidth]{c001_fullpost_new.pdf}
    \caption{Full posterior of the {\tt c001} test cosmology. True values for the parameters are shown by the solid black lines. Overall, we find that the posterior range for the HOD parameters $\log M_{\rm cut}$, $\log M_1$, $\log\sigma$, $\alpha$, $\kappa$ consistently hit the limits of our priors and are uncorrelated with our cosmological parameters. We then marginalize over these HOD parameters for the remainder of our analysis. }
    \label{fig:c001_full_post}
\end{figure}
\begin{figure}[h!]
    \centering
    \includegraphics[width=\columnwidth]{c002_fullpost_new.pdf}
    \caption{Full posterior of the {\tt c002} test cosmology. True values for the parameters are shown by the solid black lines. Overall, we find that the posterior range for the HOD parameters $\log M_{\rm cut}$, $\log M_1$, $\log\sigma$, $\alpha$, $\kappa$ consistently hit the limits of our priors and are uncorrelated with our cosmological parameters. We then marginalize over these HOD parameters for the remainder of our analysis. }
    \label{fig:c002_full_post}
\end{figure}
\begin{figure}[h!]
    \centering
    \includegraphics[width=\columnwidth]{c003_fullpost_new.pdf}
    \caption{Full posterior of the {\tt c003} test cosmology. True values for the parameters are shown by the solid black lines. Overall, we find that the posterior range for the HOD parameters $\log M_{\rm cut}$, $\log M_1$, $\log\sigma$, $\alpha$, $\kappa$ consistently hit the limits of our priors and are uncorrelated with our cosmological parameters. We then marginalize over these HOD parameters for the remainder of our analysis. }
    \label{fig:c003_full_post}
\end{figure}
\begin{figure}[h!]
    \centering
    \includegraphics[width=\columnwidth]{c004_fullpost_new.pdf}
    \caption{Full posterior of the {\tt c004} test cosmology. True values for the parameters are shown by the solid black lines. Overall, we find that the posterior range for the HOD parameters $\log M_{\rm cut}$, $\log M_1$, $\log\sigma$, $\alpha$, $\kappa$ consistently hit the limits of our priors and are uncorrelated with our cosmological parameters. We then marginalize over these HOD parameters for the remainder of our analysis. }
    \label{fig:c004_full_post}
\end{figure}
\begin{figure}[h!]
    \centering
    \includegraphics[width=\columnwidth]{c013_fullpost_new.pdf}
    \caption{Full posterior of the {\tt c013} test cosmology. True values for the parameters are shown by the solid black lines. Overall, we find that the posterior range for the HOD parameters $\log M_{\rm cut}$, $\log M_1$, $\log\sigma$, $\alpha$, $\kappa$ consistently hit the limits of our priors and are uncorrelated with our cosmological parameters. We then marginalize over these HOD parameters for the remainder of our analysis. }
    \label{fig:c013_full_post}
\end{figure}
\newpage


\bibliographystyle{JHEP}
\bibliography{Paper1_EDITS} 






    \label{fig:CMASS_posterior}



\label{lastpage}
\end{document}